\documentstyle[epsf,cite,a4wide,twocolumn,html]{article}  
\newcommand{\bfig}{\begin{figure}[bt]}
\newcommand{\Appendix}{\appendix}

\makeatletter                                                            %html
\renewcommand\section{\@startsection {section}{1}{\z@}% 		 %html
                                   {-3.5ex \@plus -1ex \@minus -.2ex}% 	 %html
                                   {2.3ex \@plus.2ex}% 			 %html
                                   {\normalfont\large\bfseries}}	 %html
\renewcommand\subsection{\@startsection{subsection}{2}{\z@}% 		 %html
                                     {-3.25ex\@plus -1ex \@minus -.2ex}% %html
                                     {1.5ex \@plus .2ex}% 		 %html
                                     {\em}}				 %html
\makeatother                                                             %html

\begin{document}

\newcommand{\be}{\begin{equation}}
\newcommand{\ee}{\end{equation}}
\newcommand{\av}[1]{\langle #1 \rangle}

\title{Surrogate time series}

\author{Thomas Schreiber and Andreas Schmitz\\
      Physics Department, University of Wuppertal,
      D--42097 Wuppertal, Germany}
%nohtml\maketitle

\date{\today\\[2cm] \parbox{\textwidth}{\normalfont\normalsize         %html
{\em Abstract}~ Before we apply nonlinear techniques, for example those
  inspired by chaos theory, to dynamical phenomena occurring in nature, it is
  necessary to first ask if the use of such advanced techniques is justified
  {\em by the data}. While many processes in nature seem very unlikely a priori
  to be linear, the possible nonlinear nature might not be evident in specific
  aspects of their dynamics. The method of surrogate data has become a very
  popular tool to address such a question. However, while it was meant to
  provide a statistically rigorous, foolproof framework, some limitations and
  caveats have shown up in its practical use. In this paper, recent efforts to
  understand the caveats, avoid the pitfalls, and to overcome some of the
  limitations, are reviewed and augmented by new material. In particular, we
  will discuss specific as well as more general approaches to constrained
  randomisation, providing a full range of examples. New algorithms will be
  introduced for unevenly sampled and multivariate data and for surrogate spike
  trains.  The main limitation, which lies in the interpretability of the test
  results, will be illustrated through instructive case studies. We will also
  discuss some implementational aspects of the realisation of these methods in
  the TISEAN software package.% 
}} %html

\maketitle %html

~\\[-2.3cm]                   %html 
\tableofcontents	      %html 
\vfill			      %html 
\eject                        %html 

\section{Introduction}
A nonlinear approach to analysing time series
data~\cite{SFI,coping,abarbook,ourbook,habil} can be motivated by two distinct
reasons.  One is intrinsic to the signal itself while the other is due to
additional knowledge we may have about the nature of the observed
phenomenon. As for the first motivation, it might be that the arsenal of linear
methods has been exploited thoroughly but all the efforts left certain
structures in the time series unaccounted for. As for the second, a system may
be known to include nonlinear components and therefore a linear description
seems unsatisfactory in the first place. Such an argument is often heard for
example in brain research --- nobody expects for example the brain to be a
linear device. In fact, there is ample evidence for nonlinearity in particular
in small assemblies of neurons.  Nevertheless, the latter reasoning is rather
dangerous. The fact that a system contains nonlinear components does not prove
that this nonlinearity is also reflected in a specific signal we measure from
that system. In particular, we do not know if it is of any practical use to go
beyond the linear approximation when analysing the signal. After all, we do not
want our data analysis to reflect our prejudice about the underlying system but
to represent a fair account of the structures that are present in the
data. Consequently, the application of nonlinear time series methods has to be
justified by establishing nonlinearity in the time series.

Suppose we had measured the signal shown in Fig.~\ref{fig:arspikes} in some
biological setting. Visual inspection immediately reveals nontrivial structure
in the serial correlations. The data fails a test for Gaussianity, thus ruling
out a Gaussian linear stochastic process as its source.  Depending on the
assumptions we are willing to make on the underlying process, we might suggest
different origins for the observed strong ``spikyness'' of the dynamics.
Superficially, low dimensional chaos seems unlikely due to the strong
fluctuations, but maybe high dimensional dynamics? A large collection of
neurons could intermittently synchronise to give rise to the burst episodes. In
fact, certain artificial neural network models show qualitatively similar
dynamics.  The least interesting explanation, however, would be that all the
spikyness comes from a distortion by the measurement procedure and all the
serial correlations are due to linear stochastic dynamics. Occam's razor tells
us that we should be able to rule out such a simple explanation before we
venture to construct more complicated models.

Surrogate data testing attempts to find the least interesting explanation
that cannot be ruled out based on the data. In the above example, the data
shown in Fig.~\ref{fig:arspikes}, this would be the hypothesis that the data
has been generated by a stationary Gaussian linear stochastic process
(equivalently, an {\em autoregressive moving average} or ARMA process) that is
observed through an invertible, static, but possible nonlinear observation
function:
\be
   s_n=s(x_n),\quad \{x_n\}: \mbox{ARMA}(M,N)
\,.\ee 
Neither the order $M,N$, the ARMA coefficients, nor the function $s(\cdot)$ are
assumed to be known. Without explicitly modeling these parameters, we still
know that such a process would show characteristic linear correlations
(reflecting the ARMA structure) and a characteristic single time probability
distribution (reflecting the action of $s(\cdot)$ on the original Gaussian
distribution). Figure~\ref{fig:arspikes_surr} shows a surrogate time series
that is designed to have exactly these properties in common with the data but
to be as random as possible otherwise. By a proper statistical test we can now
look for additional structure that is present in the data but not in the
surrogates.

\bfig%
\centerline{% 
   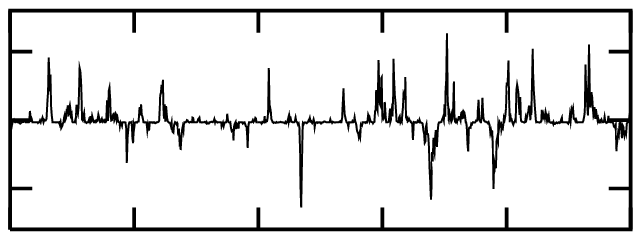% 
}
\caption[]{\label{fig:arspikes}
   A time series showing characteristic bursts.}
\end{figure}

\bfig%
\centerline{% 
   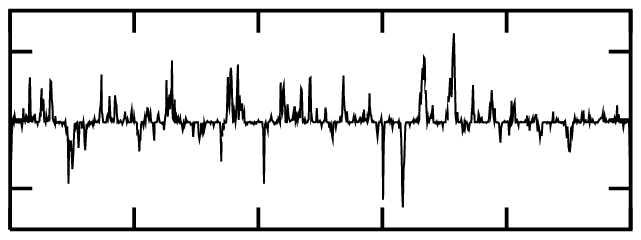% 
}
\caption[]{\label{fig:arspikes_surr}
   A surrogate time series that has the same single time probability
   distribution and the same autocorrelation function as the sequence in
   Fig.~\ref{fig:arspikes}. The bursts are fully explained by these two
   properties.}
\end{figure}

In the case of the time series in Fig.~\ref{fig:arspikes}, there is no
additional structure since it has been generated by the rule
\be
   s_n=\alpha x_n^3,\quad x_n=0.9x_{n-1} + \eta_n
\ee
where $\{\eta_n\}$ are Gaussian independent increments and $\alpha$ is chosen
so that the data have unit variance.% 
\footnote{
   In order to simplify the notation in mathematical derivations, we will
   assume throughout this paper that the mean of each time series has been
   subtracted and it has been rescaled to unit variance. Nevertheless, we will
   often transform back to the original experimental units when displaying
   results graphically.}
This means that the strong nonlinearity that generates the bursts is due to the
distorted measurement that enhances ordinary fluctuations, generated by linear
stochastic dynamics.

In order to systematically exclude simple explanations for time series
observations, this paper will discuss formal statistical tests for
nonlinearity. We will formulate suitable null hypotheses for the underlying
process or for the observed structures themselves. In the former case, null
hypotheses will be extensions of the statement that the data were generated by
a Gaussian linear stochastic processes. The latter situation may occur when it
is difficult to properly define a class of possibly underlying processes but we
want to check if a particular set of observables gives a complete account of
the statistics of the data. We will attempt to reject a null hypothesis by
comparing the value of a nonlinear parameter taken on by the data with its
probability distribution.  Since only exceptional cases allow for the exact or
asymptotic derivation of this distribution unless strong additional assumptions
are made, we have to estimate it by a Monte Carlo resampling technique. This
procedure is known in the nonlinear time series literature as the method of
{\em surrogate data}, see Refs.~\cite{theiler1,theiler-sfi,fields}. Most of the
body of this paper will be concerned with the problem of generating an
appropriate Monte Carlo sample for a given null hypothesis.

We will also dwell on the proper interpretation of the outcome of such a test.
Formally speaking, this is totally straightforward: A rejection at a given
significance level means that if the null hypothesis is true, there is certain
small probability to still see the structure we detected.  Non-rejection means
even less: either the null hypothesis is true, or the discriminating statistics
we are using fails to have power against the alternative realised in the
data. However, one is often tempted to go beyond this simple reasoning and
speculate either on the nature of the nonlinearity or non-stationarity that
lead to the rejection, or on the reason for the failure to reject.

Since the actual quantification of nonlinearity turns out to be the easiest ---
or in any case the least dangerous --- part of the problem, we will discuss it
first.  In principle, any nonlinear parameter can be employed for this purpose.
They may however differ dramatically in their ability to detect different kinds
of structures. Unfortunately, selecting the most suitable parameter has to be
done without making use of the data since that would render the test incorrect:
If the measure of nonlinearity has been optimised formally or informally with
respect to the data, a fair comparison with surrogates is no longer
possible. Only information that is shared by data and surrogates, that is, for
example, linear correlations, may be considered for guidance.  If multiple data
sets are available, one could use some sequences for the selection of the
nonlinearity parameter and others for the actual test.  Otherwise, it is
advantageous to use one of the parameter free methods that can be set up with
very little detailed knowledge of the data.

Since we want to advocate to routinely use a nonlinearity test whenever
nonlinear methods are planned to be applied, we feel that it is important to
make a practical implementation of such a test easily accessible. Therefore,
one branch of the TISEAN free software package~\cite{tisean} is devoted to
surrogate data testing. Appendix~\ref{app:A} will discuss the implementational
aspects necessary to understand what the programs in the package do.

\section{Detecting weak nonlinearity}
Many quantities have been discussed in the literature that can be used to
characterise nonlinear time series. For the purpose of nonlinearity testing we
need such quantities that are particular powerful in discriminating linear
dynamics and weakly nonlinear signatures --- strong nonlinearity is usually
more easily detectable. An important objective criterion that can be used to
guide the preferred choice is the discrimination {\em power} of the resulting
test. It is defined as the probability that the null hypothesis is rejected
when it is indeed false.  It will obviously depend on how and how strongly the
data actually deviates from the null hypothesis.

\subsection{Higher order statistics}
Traditional measures of nonlinearity are derived from generalisations of the
two-point auto-covariance function or the power spectrum. The use of higher
order cumulants as well as bi- and multi-spectra is discussed for example in
Ref.~\cite{BI}.  One particularly useful third order quantity% 
\footnote{We have omitted the commonly used normalisation to second moments
   since throughout this paper, time series and their surrogates will have the
   same second order properties and identical pre-factors do not enter the
   tests.} 
is 
\be\label{eq:skew} 
    \phi^{\rm rev}(\tau) = {1\over N-\tau}\sum_{n=\tau+1}^N (s_n-s_{n-\tau})^3 
\,,\ee
since it measures the asymmetry of a series under time reversal. (Remember that
the statistics of linear stochastic processes is always symmetric under time
reversal. This can be most easily seen when the statistical properties are
given by the power spectrum which contains no information about the direction
of time.) Time reversibility as a criterion for discriminating time series
is discussed in detail in Ref.~\cite{diks2}, where, however, a different
statistic is used to quantify it. The concept itself is quite folklore and has
been used for example in Refs.~\cite{theiler1,Timmer1}.

Time irreversibility can be a strong signature of nonlinearity. Let us point
out, however, that it does not imply a dynamical origin of the nonlinearity.
We will later (Sec.~\ref{sec:rev}) give an example of time asymmetry
generated by a measurement function involving a nonlinear time average.

\subsection{Phase space observables}
When a nonlinearity test is performed with the question in mind if nonlinear
deterministic modeling of the signal may be useful, it seems most appropriate
to use a test statistic that is related to a nonlinear deterministic approach.
We have to keep in mind, however, that a positive test result only indicates
nonlinearity, not necessarily determinism. Since nonlinearity tests are usually
performed on data sets which do not show unambiguous signatures of
low-dimensional determinism (like clear scaling over several orders of
magnitude), one cannot simply estimate one of the quantitative indicators of
chaos, like the fractal dimension or the Lyapunov exponent. The formal answer
would almost always be that both are probably infinite. Still, some useful test
statistics are at least inspired by these quantities. Usually, some effective
value at a finite length scale has to be computed without establishing scaling
region or attempting to approximate the proper limits.

In order to define an observable in $m$--dimensional phase space, we first have
to reconstruct that space from a scalar time series, for example by the method
of delays:
\be\label{eq:delay}
   {\bf s}_n=(s_{n-(m-1)\tau}, s_{n-(m-2)\tau}, \ldots, s_n)
\,.\ee
One of the more robust choices of phase space observable is a nonlinear
prediction error with respect to a locally constant predictor
$F$ that can be defined by
\be\label{eq:error}
   \gamma (m,\tau,\epsilon) =
      \left({1\over N} \sum [s_{n+1}-F({\bf s}_n)]^2\right)^{1/2} 
\,.\ee
The prediction over one time step is performed by averaging over the future
values of all neighbouring delay vectors closer than $\epsilon$ in $m$
dimensions.   

We have to consider the limiting case that the deterministic signature to be
detected is weak. In that case, the major limiting factor for the performance
of a statistical indicator is its variance since possible differences between
two samples may be hidden among the statistical fluctuations. In
Ref.~\cite{power}, a number of popular measures of nonlinearity are compared
quantitatively. The results can be summarised by stating that in the presence
of time-reversal asymmetry, the particular quantity Eq.(\ref{eq:skew}) that
derives from the three-point autocorrelation function gives very reliable
results. However, many nonlinear evolution equations produce little or no
time-reversal asymmetry in the statistical properties of the signal. In these
cases, simple measures like a prediction error of a locally constant phase
space predictor, Eq.(\ref{eq:error}), performed best.  It was found to be
advantageous to choose embedding and other parameters in order to obtain a
quantity that has a small spread of values for different realisations of the
same process, even if at these parameters no valid embedding could be expected.

Of course, prediction errors are not the only class of nonlinearity measures
that has been optimised for robustness. Notable other examples are
coarse-grained redundancies~\cite{milan2,pompe,pt}, and, at an even higher
level of coarse-graining, symbolic methods~\cite{hao}. The very popular method
of {\em false nearest neighbours}~\cite{FNN} can be easily modified to yield a
scalar quantity suitable for nonlinearity testing. The same is true for the
concept of {\em unstable periodic orbits} (UPOs)~\cite{PM,soso}.

\section{Surrogate data testing}
All of the measures of nonlinearity mentioned above share a common property.
Their probability distribution on finite data sets is not known analytically --
except maybe when strong additional assumptions about the data are made. Some
authors have tried to give error bars for measures like predictabilities (e.g.
Barahona and Poon~\cite{volterra}) or averages of pointwise dimensions (e.g.
Skinner et al.~\cite{skinner}) based on the observation that these quantities
are averages (mean values or medians) of many individual terms, in which case
the variance (or quartile points) of the individual values yield an error
estimate. This reasoning is however only valid if the individual terms are
independent, which is usually not the case for time series data. In fact, it is
found empirically that nonlinearity measures often do not even follow a
Gaussian distribution.  Also the standard error given by
Roulston~\cite{roulston} for the mutual information is fully correct only for
uniformly distributed data. His derivation assumes a smooth rescaling to
uniformity. In practice, however, we have to rescale either to {\em exact}
uniformity or by rank-ordering uniform variates. Both transformations are in
general non-smooth and introduce a bias in the joint probabilities.  In view of
the serious difficulties encountered when deriving confidence limits or
probability distributions of nonlinear statistics with analytical methods, it
is highly preferable to use a Monte Carlo resampling technique for this
purpose.

\subsection{Typical vs. constrained realisations}
Traditional bootstrap methods use explicit model equations that have to be
extracted from the data and are then run to produce Monte Carlo samples.
This {\em typical realisations} approach can be very powerful for the
computation of confidence intervals, provided the model equations can be
extracted successfully. The latter requirement is very delicate. Ambiguities in
selecting the proper model class and order, as well as the parameter estimation
problem have to be addressed. Whenever the null hypothesis involves an unknown
{\em function} (rather than just a few parameters) these problems become
profound. A recent example of a {\em typical realisations} approach to creating
surrogates in the dynamical systems context is given by Ref.~\cite{witt}.
There, a Markov model is fitted to a coarse-grained dynamics obtained by
binning the two dimensional delay vector distribution of a time series.
Then, essentially the transfer matrix is iterated to yield surrogate
sequences. We will offer some discussion of that work later in
Sec.~\ref{sec:interpret}. 

As discussed by Theiler and Prichard~\cite{tp}, the
alternative approach of {\em constrained realisations} is more suitable for the
purpose of hypothesis testing we are interested in here. It avoids the fitting
of model equations by directly imposing the desired structures onto the
randomised time series.  However, the choice of possible null hypothesis is
limited by the difficulty of  imposing arbitrary structures on otherwise random
sequences. In the following, we will discuss a number of null hypotheses and
algorithms to provide the adequately constrained realisations. The most general
method to generate constrained randomisations of time series~\cite{anneal} is 
described in Sec.~\ref{sec:anneal}.

Consider as a toy example the null hypothesis that the data consists of
independent draws from a fixed probability distribution. Surrogate time series
can be simply obtained by randomly shuffling the measured data. If we find
significantly different serial correlations in the data and the shuffles, we
can reject the hypothesis of independence. Constrained realisations are
obtained by creating permutations {\em without replacement}. The surrogates are
constrained to take on exactly the same values as the data, just in random
temporal order. We could also have used the data to infer the probability
distribution and drawn new time series from it. These permutations {\em with
replacement} would then be what we called typical realisations.

Obviously, independence is not an interesting null hypothesis for most time
series problems. It becomes relevant when the residual errors of a time series
model are evaluated. For example in the BDS test for
nonlinearity~\cite{brockpaper1}, an ARMA model is fitted to the data. If the
data are linear, then the residuals are expected to be independent.
It has been pointed out, however, that the resulting test is not particularly
powerful for chaotic data~\cite{bleach}.

\subsection{The null hypothesis: model class vs. properties}
From the bootstrap literature we are used to defining null hypothesis for time
series in terms of a class of processes that is assumed to contain the specific
process that generated the data. For most of the literature on surrogate data,
this situation hasn't changed. One very common null hypothesis goes back to
Theiler and coworkers~\cite{theiler1} and states that the data have been
generated by a Gaussian linear stochastic process with constant coefficients.
Constrained realisations are created by requiring that the surrogate time
series have the same Fourier amplitudes as the data. We can clearly see in this
example that what is needed for the constrained realisations approach is a set
of observable properties that is known to fully specify the process. The
process itself is not reconstructed. But this example is also exceptional.  We
know that the class of processes defined by the null hypothesis is fully
parametrised by the set of ARMA$(M,N)$ models (autoregressive moving average,
see Eq.(\ref{eq:arma}) below).  If we allow for arbitrary orders $M$ and $N$,
there is a one-to-one correspondence between the ARMA coefficients and the
power spectrum. The power spectrum is here estimated by the Fourier
amplitudes. The Wiener--Khinchin theorem relates it to the autocorrelation
function by a simple Fourier transformation. Consequently, specifying either
the class of processes or the set of constraints are two ways to achieve the
same goal.  The only generalisation of this favourable situation that has been
found so far is the null hypothesis that the ARMA output may have been observed
by a static, invertible measurement function. In that case, constraining the
single time probability distribution and the Fourier amplitudes is sufficient.

If we want to go beyond this hypothesis, all we can do in general is to specify
the set of constraints we will impose. We cannot usually say which class of
processes this choice corresponds to. We will have to be content with
statements that a given set of statistical parameters exhaustively describes
the statistical properties of a signal. Hypotheses in terms of a model class
are usually more informative but specifying sets of observables gives us much
more flexibility.

\subsection{Test design}
Before we go into detail about the generation of surrogate samples, let us
outline how an actual test can be carried out.  Many examples are known of
nonlinearity measures that aren't even approximately normally distributed. It
has therefore been advocated since the early days~\cite{theiler1} to use robust
statistics rather than parametric methods for the actual statistical test. In
other words, we discourage the common practice to represent the distribution of
the nonlinearity measure by an error bar and deriving the significance from the
number of ``sigmas'' the data lies outside these bounds. Such a reasoning
implicitly assumes a Gaussian distribution.

Instead, we follow Theiler et al.~\cite{theiler1} by using a rank--order test.
First, we select a residual probability $\alpha$ of a false rejection,
corresponding to a level of significance $(1-\alpha)\times 100\% $.  Then, for a
one--sided test (e.g. looking for {\em small} prediction errors only), we
generate $M=1/\alpha-1$ surrogate sequences. Thus, including the data itself,
we have $1/\alpha$ sets. Therefore, the probability that the data by
coincidence has the smallest, say, prediction error is exactly $\alpha$, as
desired.  For a two--sided test (e.g. for time asymmetry which can go both
ways), we would generate $M=2/\alpha-1$ surrogates, resulting in a probability
$\alpha$ that the data gives {\em either} the smallest {\em or} the largest
value. 

For a minimal significance requirement of 95\% , we thus need at least 19 or 39
surrogate time series for one-- and two--sided tests, respectively. The
conditions for rank based tests with more samples can be easily worked out.
Using more surrogates can increase the discrimination power.

\section{Fourier based surrogates}\label{sec:fourier}
In this section, we will discuss a hierarchy of null hypotheses and the
issues that arise when creating the corresponding surrogate data. The simpler
cases are discussed first in order to illustrate the reasoning. If we have
found serial correlations in a time series, that is, rejected the null
hypothesis of independence, we may ask of what nature these correlations are.
The simplest possibility is to explain the observed structures by
linear two-point autocorrelations. A corresponding null hypothesis is that the
data have been generated by some linear stochastic process with Gaussian
increments. The most general univariate linear process is given by
\be\label{eq:arma}
   s_n=\sum_{i=1}^M a_is_{n-i} + \sum_{i=0}^N b_i\eta_{n-i}
\,,\ee
where $\{\eta_n\}$ are Gaussian uncorrelated random increments.  The
statistical test is complicated by the fact that we do not want to test against
one particular linear process only (one specific choice of the $a_i$ and
$b_i$), but against a whole class of processes. This is called a {\em
composite} null hypothesis. The unknown values $a_i$ and $b_i$ are sometimes
referred to as {\em nuisance parameters}. There are basically three directions
we can take in this situation. First, we could try to make the discriminating
statistic independent of the nuisance parameters. This approach has not been
demonstrated to be viable for any but some very simple statistics. Second, we
could determine which linear model is most likely realised in the data by a fit
for the coefficients $a_i$ and $b_i$, and then test against the hypothesis that
the data has been generated by this particular model. Surrogates are simply
created by running the fitted model. This {\em typical realisations} approach
is the common choice in the bootstrap literature, see e.g.  the classical book
by Efron~\cite{efron}. The main drawback is that we cannot recover the {\em
true} underlying process by any fit procedure. Apart from problems associated
with the choice of the correct model orders $M$ and $N$, the data is by
construction a very likely realisation of the fitted process.  Other
realisations will fluctuate {\em around} the data which induces a bias against
the rejection of the null hypothesis. This issue is discussed thoroughly in
Ref.~\cite{fields}, where also a calibration scheme is proposed.

The most attractive approach to testing for a composite null hypothesis seems
to be to create {\em constrained realisations}~\cite{tp}. Here it is useful to
think of the measurable properties of the time series rather than its
underlying model equations. The null hypothesis of an underlying Gaussian
linear stochastic process can also be formulated by stating that all structure
to be found in a time series is exhausted by computing first and second order
quantities, the mean, the variance and the auto-covariance function.  This
means that a randomised sample can be obtained by creating sequences with the
same second order properties as the measured data, but which are otherwise
random. When the linear properties are specified by the squared amplitudes of
the (discrete) Fourier transform
\be\label{eq:pgram}
    |S_k|^2 = \left| {1\over \sqrt{N}} 
          \sum_{n=0}^{N-1} s_n e^{i2\pi kn/N} \right|^2
\,,\ee
that is, the periodogram estimator of the power spectrum, surrogate time series
$\{\overline{s}_n\}$ are readily created by multiplying the Fourier transform
of the data by random phases and then transforming back to the time domain:
\be\label{eq:ftsurro}
   \overline{s}_n = {1\over \sqrt{N}} 
          \sum_{k=0}^{N-1} e^{i\alpha_k} |S_k| \,e^{-i2\pi kn/N}
\,,\ee
where $0\le \alpha_k<2\pi$ are independent uniform random numbers.

\subsection{Rescaled Gaussian linear process}
The two null hypotheses discussed so far (independent random numbers and
Gaussian linear processes) are not what we want to test against in most
realistic situations. In particular, the most obvious deviation from the
Gaussian linear process is usually that the data do not follow a Gaussian
single time probability distribution. This is quite obvious for data obtained
by measuring intervals between events, e.g. heart beats since intervals are
strictly positive. There is however a simple generalisation of the null
hypothesis that explains deviations from the normal distribution by the action
of an invertible, static measurement function: 
\be\label{eq:distort}
   s_n=s(x_n),\quad x_n=\sum_{i=1}^M a_ix_{n-i} + \sum_{i=0}^N b_i\eta_{n-i}
\,.\ee 
We want to regard a time series from such a process as essentially linear since
the only nonlinearity is contained in the --- in principle invertible ---
measurement function $s(\cdot)$.

Let us mention right away that the restriction that $s(\cdot)$ must be
invertible is quite severe and often undesired. The reason why we have to
impose it is that otherwise we couldn't give a complete specification of the
process in terms of observables and constraints. The problem is further
illustrated in Sec.~\ref{sec:rev} below.

The most common method to create surrogate data sets for this null hypothesis
essentially attempts to invert $s(\cdot)$ by rescaling the time series
$\{s_n\}$ to conform with a Gaussian distribution. The rescaled version is
then phase randomised (conserving Gaussianity on average) and the result is
rescaled to the empirical distribution of $\{s_n\}$. The rescaling is done by
simple rank ordering. Suppose we want to rescale the sequence $\{s_n\}$ so that
the rescaled sequence $\{r_n\}$ takes on the same values as some reference
sequence $\{g_n\}$ (e.g. draws from a Gaussian distribution). Let $\{g_n\}$ be
sorted in ascending order and $\mbox{rank}(s_n)$ denote the ascending rank of
$s_n$, e.g. $\mbox{rank}(s_n)=3$ if $s_n$ is the 3rd smallest element of 
$\{s_n\}$. Then the rescaled sequence is given by
\be\label{eq:rank}
   r_n=g_{{\rm\scriptsize rank}(s_n)},\quad n=1,\ldots,N
\,.\ee
The {\em amplitude adjusted Fourier transform} (AAFT) method has been
originally proposed by Theiler et al.~\cite{theiler1}. It results in a
correct test when $N$ is large, the correlation in the data is not too strong
and $s(\cdot)$ is close to the identity. Otherwise, there is a certain bias
towards a too flat spectrum, to be discussed in the following section.

\subsection{Flatness bias of AAFT surrogates}
It is argued in Ref.~\cite{surrowe} that for short and strongly correlated
sequences, the AAFT algorithm can yield an incorrect test since it introduces a
bias towards a slightly flatter spectrum. In Fig.~\ref{fig:flat} we see power
spectral estimates of a clinical data set and of 19 AAFT surrogates.  The data
is taken from data set B of the Santa Fe Institute time series
contest~\cite{gold}. It consists of 4096 samples of the breath rate of a
patient with sleep apnoea. The sampling interval is 0.5~seconds. The
discrepancy of the spectra is significant. A bias towards a white spectrum is
noted: power is taken away from the main peak to enhance the low and high
frequencies.

\bfig%
\centerline{% 
   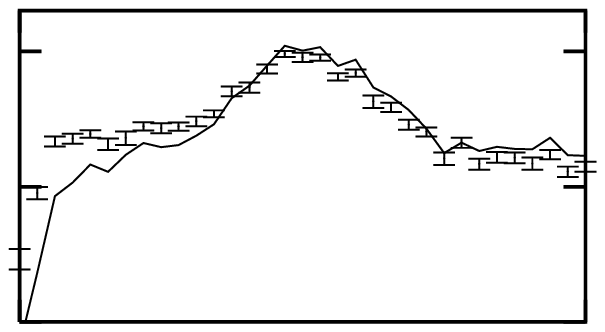% 
}
\caption[]{\label{fig:flat}
   Discrepancy of the power spectra of human breath rate data (solid
   line) and 19 AAFT surrogates (dashed lines). Here the power spectra have
   been  computed with a square window of length~64.}
\end{figure}

Heuristically, the flatness bias can be understood as follows. Amplitude
adjustment attempts to invert the unknown measurement function $s(\cdot)$
empirically. The estimate $\hat{s}^{-1}(\cdot)$ of the inverse obtained by the
rescaling of a finite sample to values drawn from a Gaussian distribution is
expected to be consistent but it is not exact for finite $N$. The sampling
fluctuations of $\delta_n=\hat{s}^{-1}(s_n)-s^{-1}(s_n)$ will be essentially
independent of $n$ and thus spectrally white. Consequently, Gaussian scaling
amounts to adding a white component to the spectrum, which therefore tends to
become flatter under the procedure. Since such a bias can lead to spurious
results, surrogates have to be refined before a test can be performed.

\subsection{Iteratively refined surrogates}\label{sec:iterative}
In Ref.~\cite{surrowe}, we propose a method which iteratively corrects
deviations in spectrum and distribution from the goal set by the measured
data. In an alternating fashion, the surrogate is filtered towards the correct
Fourier amplitudes and rank-ordered to the correct distribution.

Let $\{|S_k|^2\}$ be the Fourier amplitudes, Eq.(\ref{eq:pgram}), of the
data and $\{c_k\}$ a copy of the data sorted by magnitude in ascending order.
At each iteration stage $(i)$, we have a sequence $\{\overline{r}_n^{(i)}\}$
that has the correct distribution (coincides with $\{c_k\}$ when sorted), and a
sequence $\{\overline{s}_n^{(i)}\}$ that has the correct Fourier amplitudes
given by $\{|S_k|^2\}$.  One can start with $\{\overline{r}_n^{(0)}\}$
being either an AAFT surrogate, or simply a random shuffle of the data.  

The step $\overline{r}_n^{(i)}\to \overline{s}_n^{(i)}$ is a very crude
``filter'' in the Fourier domain: The Fourier amplitudes are simply {\em
replaced} by the desired ones. First, take the (discrete) Fourier transform of
$\{\overline{r}_n^{(i)}\}$:
\be
    \overline{R}_k^{(i)} = 
       {1\over \sqrt{N}} \sum_{n=0}^{N-1} \overline{r}_n e^{i2\pi kn/N}
\,.\ee
Then transform back, replacing the actual amplitudes by the desired ones, but
keeping the phases $e^{i\psi_k^{(i)}}=\overline{R}_k^{(i)} / 
|\overline{R}_k^{(i)}|$:
\be\label{eq:step1}
    \overline{s}_n^{(i)} = {1\over \sqrt{N}} 
       \sum_{k=0}^{N-1} 
         e^{i\psi_k^{(i)}} |S_k|\,
         e^{-i2\pi kn/N}
\,.\ee
The step $\overline{s}_n^{(i)}\to \overline{r}_n^{(i+1)}$ proceeds by rank
ordering:
\be\label{eq:step2}
   \overline{r}_n^{(i+1)}=c_{\mbox{\scriptsize rank}(\overline{s}_n^{(i)})}
\,.\ee
It can be heuristically understood that the iteration scheme is attracted to a
fixed point $\overline{r}_n^{(i+1)}=\overline{r}_n^{(i)}$ for large
$(i)$. Since the minimal possible change equals to the smallest nonzero
difference $c_n-c_{n-1}$ and is therefore finite for finite $N$, the fixed
point is reached after a finite number of iterations. The remaining discrepancy
between $\overline{r}_n^{(\infty)}$ and $\overline{s}_n^{(\infty)}$ can be
taken as a measure of the accuracy of the method. Whether the residual bias in
$\overline{r}_n^{(\infty)}$ or $\overline{s}_n^{(\infty)}$ is more tolerable
depends on the data and the nonlinearity measure to be used. For coarsely
digitised data,% 
\footnote{ 
   Formally, digitisation is a non-invertible, nonlinear measurement and thus
   not included in the null hypothesis. Constraining the surrogates to take
   exactly the same (discrete) values as the data seems to be reasonably safe,
   though. Since for that case we haven't seen any dubious rejections due to
   discretisation, we didn't discuss this issue as a serious caveat. This
   decision may of course prove premature.}
deviations from the discrete distribution can lead to spurious results
whence $\overline{r}_n^{(\infty)}$ is the safer choice. If linear correlations
are dominant, $\overline{s}_n^{(\infty)}$ can be more suitable.

The final accuracy that can be reached depends on the size and structure of the
data and is generally sufficient for hypothesis testing. In all the cases we
have studied so far, we have observed a substantial improvement over the
standard AAFT approach. Convergence properties are also discussed
in~\cite{surrowe}. In Sec.~\ref{sec:accuracy} below, we will say more about the
remaining inaccuracies.

\subsection{Example: Southern oscillation index}

As an illustration let us perform a statistical test for nonlinearity on a
monthly time series of the Southern Oscillation Index (SOI) from 1866 to 1994
(1560 samples).  For a reference on analysis of Southern Oscillation data
see Graham et al.~\cite{Graham87}. Since a discussion of this 
climatic phenomenon is not relevant to the issue at hand, let us just consider
the time series as an isolated data item. Our null hypothesis is that the data
is adequately described by its single time probability distribution and its
power spectrum.  This corresponds to the assumption that an autoregressive
moving average (ARMA) process is generating a sequence that is measured through
a static monotonic, possibly nonlinear observation function.

For a test at the 99\%  level of significance ($\alpha=0.01$), we generate a
collection of $1/\alpha-1=99$ surrogate time series which share the single time
sample probability distribution and the periodogram estimator with the
data. This is carried out using the iterative method described in
Sec.~\ref{sec:iterative} above (see also Ref.~\cite{surrowe}).
Figure~\ref{fig:soi} shows the data together with one of the 99 surrogates.

As a discriminating statistics we use a locally constant predictor in embedding
space, using three dimensional delay coordinates at a delay time of one
month. Neighbourhoods were selected at 0.2 times the rms amplitude of the
data. The test is set up in such a way that the null hypothesis may be rejected
when the prediction error is smaller for the data than for all of the 99
surrogates. But, as we can see in Fig.~\ref{fig:soipred}, this is not the
case. Predictability is not significantly reduced by destroying possible
nonlinear structure.  This negative result can mean several things. The
prediction error statistics may just not have any power to detect the kind of
nonlinearity present.  Alternatively, the underlying process may be linear and
the null hypothesis true. It could also be, and this seems the most likely
option after all we know about the equations governing climate phenomena,
that the process is nonlinear but the single time series at this sampling
covers such a poor fraction of the rich dynamics that it must appear linear
stochastic to the analysis.

\bfig%
\centerline{% 
   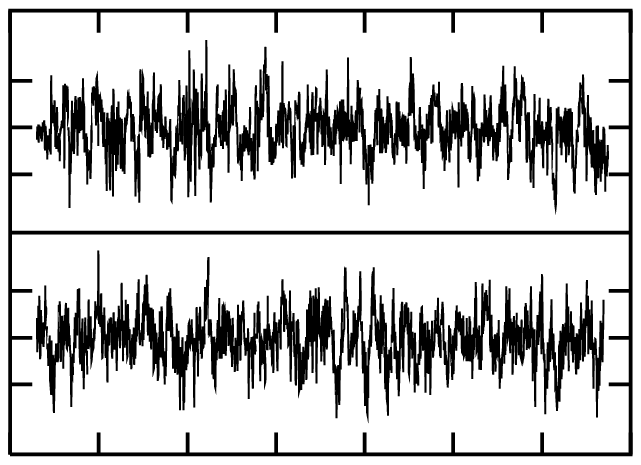% 
}
\caption[]{\label{fig:soi} 
   Monthly values of the Southern Oscillation Index (SOI) from 1866 to 1994
   (upper trace) and a surrogate time series exhibiting the same
   auto-covariance function (lower trace). All linear properties of the
   fluctuations and oscillations are the same between both tracings. However,
   any possible nonlinear structure except for a static rescaling of the data
   is destroyed in the lower tracing by the randomisation procedure.}
\end{figure}

\bfig%
\centerline{% 
   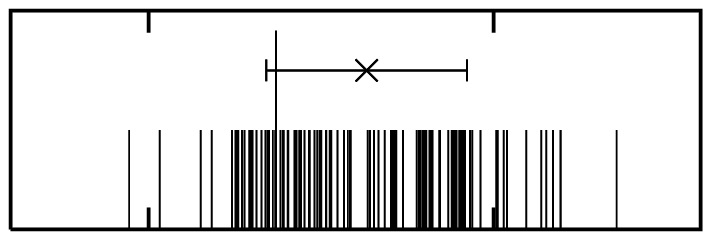% 
}
\caption[]{\label{fig:soipred}
   Nonlinear prediction error measured for the SOI data set (see
   Fig.~\ref{fig:soi}) and 99 surrogates. The value for the original data is
   plotted with a longer impulse. The mean and standard deviation of the
   statistic obtained from the surrogates is also represented by an error
   bar. It is evident that the data is not singled out by this property and we
   are unable to reject the null hypothesis of a linear stochastic stationary
   process, possibly rescaled by a nonlinear measurement function.}
\end{figure}

Of course, our test has been carried out disregarding any knowledge of the SOI
situation. It is very likely that more informed measures of nonlinearity may be
more successful in detecting structure. We would like to point out, however,
that if such information is derived from the same data, or literature published
on it, a bias is likely to occur. Similarly to the situation of multiple tests
on the same sample, the level of significance has to be adjusted properly.
Otherwise, if many people try, someone will eventually, and maybe accidentally,
find a measure that indicates nonlinear structure.

\subsection{Periodicity artefacts}\label{sec:period}

\bfig%
\centerline{% 
   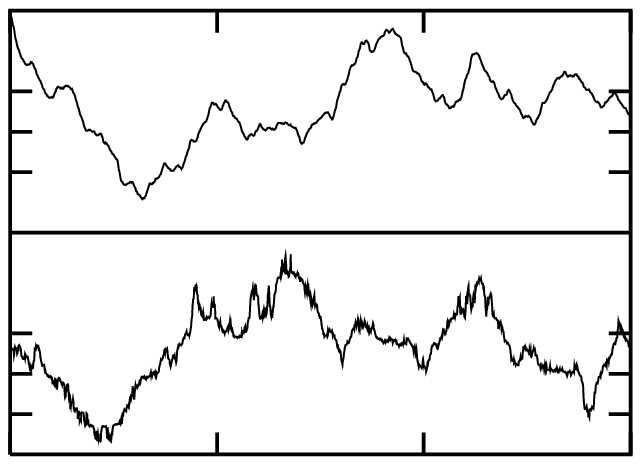% 
}
\caption[]{\label{fig:end}
   Effect of end point mismatch on Fourier based surrogates. Upper trace:
   1500 iterates of $s_n=1.9s_{n-1} - 0.9001s_{n-2} + \eta_n$. Lower trace:
   a surrogate sequence with the same Fourier amplitudes. Observe the
   additional ``crinkliness'' of the surrogate.}
\end{figure}

The randomisation schemes discussed so far all base the quantification of
linear correlations on the Fourier amplitudes of the data. Unfortunately, this
is not exactly what we want. Remember that the autocorrelation structure given
by
\be\label{eq:autocor} 
   C(\tau)={1\over N-\tau}\sum_{n=\tau+1}^{N} s_n s_{n-\tau} 
\ee
corresponds to the Fourier amplitudes {\em only} if the time series is one
period of a sequence that repeats itself every $N$ time steps. This is, however,
not what we believe to be the case. Neither is it compatible with the null
hypothesis. Conserving the Fourier amplitudes of the data means that the {\em
periodic} auto-covariance function
\be\label{eq:cp}
   C_p(\tau) = {1\over N}
      \sum_{n=1}^{N} s_n s_{{\rm\scriptsize mod}(n-\tau-1,N)+1}
\ee
is reproduced, rather than $C(\tau)$. This seemingly harmless difference can
lead to serious artefacts in the surrogates, and, consequently, spurious
rejections in a test.  In particular, any mismatch between the beginning and
the end of a time series poses problems, as discussed e.g. in
Ref.~\cite{theiler-sfi}. In spectral estimation, problems caused by edge
effects are dealt with by windowing and zero padding. None of these techniques
have been successfully implemented for the phase randomisation of surrogates
since they destroy the invertibility of the transform.

Let us illustrate the artefact generated by an end point mismatch with an
example. In order to generate an effect that is large enough to be detected
visually, consider 1500 iterates of the almost unstable AR(2) process,
$s_n=1.9s_{n-1} - 0.9001s_{n-2} + \eta_n$ (upper trace of Fig.~\ref{fig:end}).
The sequence is highly correlated and there is a rather big difference between
the first and the last points. Upon periodic continuation, we see a jump
between $s_{1500}$ and $s_1$. Such a jump has spectral power at all
frequencies but with delicately tuned phases.  In surrogate time series
conserving the Fourier amplitudes, the phases are randomised and the spectral
content of the jump is spread in time. In the surrogate sequence shown as the
lower trace in Fig.~\ref{fig:end}, the additional spectral power is mainly
visible as a high frequency component.  It is quite clear that the difference
between the data and such surrogates will be easily been picked up by, say, a
nonlinear predictor, and can lead to spurious rejections of the null
hypothesis.

\bfig%
\centerline{% 
   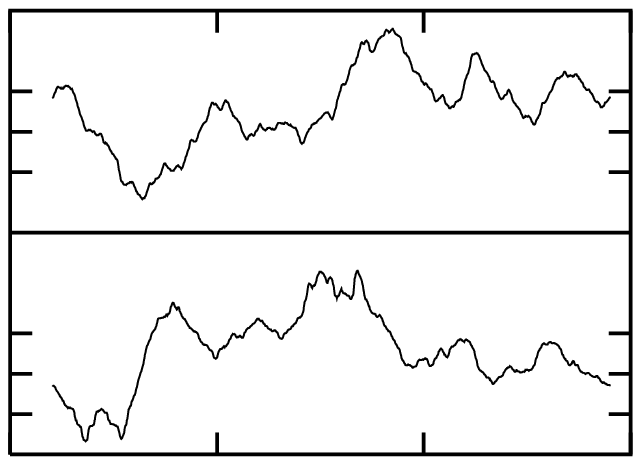% 
}
\caption[]{\label{fig:endno}
   Repair of end point mismatch by selecting a sub-sequence of length 1350 of
   the signal shown in Fig.~\ref{fig:end} that has an almost perfect match of 
   end points. The surrogate shows no spurious high frequency structure.}
\end{figure}

The problem of non-matching ends can often be overcome by choosing a
sub-interval of the recording such that the end points do match as closely as
possible~\cite{t_neuro}. The possibly remaining finite phase slip at the
matching points usually is of lesser importance. It can become dominant,
though, if the signal is otherwise rather smooth.  As a systematic strategy,
let us propose to measure the end point mismatch by
\be\label{eq:mismatch}
   \gamma_{\rm\scriptsize jump}
      ={(s_1-s_N)^2 \over \sum_{n=1}^N (s_n-\av{s})^2}
\ee
and the mismatch in the first derivative by
\be\label{eq:slip}
   \gamma_{\rm\scriptsize slip}
      ={[(s_2-s_1)-(s_N-s_{N-1})]^2 \over \sum_{n=1}^N (s_n-\av{s})^2}
\,.\ee
The fractions $\gamma_{\rm\scriptsize jump}$ and $\gamma_{\rm\scriptsize slip}$
give the contributions to the total power of the series of the mismatch of the
end points and the first derivatives, respectively. For the series shown in
Fig.~\ref{fig:end}, $\gamma_{\rm\scriptsize jump}=0.45\%$ and the end effect
dominates the high frequency end of the spectrum.  By systematically going
through shorter and shorter sub-sequences of the data, we find that a segment
of 1350 points starting at sample 102 yields $\gamma_{\rm\scriptsize
jump}=10^{-5}\%$ or an almost perfect match.  That sequence is shown as the
upper trace of Fig.~\ref{fig:endno}, together with a surrogate (lower
trace). The spurious ``crinkliness'' is removed.

In practical situations, the matching of end points is a simple and mostly
sufficient precaution that should not be neglected. Let us mention that the SOI
data discussed before is rather well behaved with little end-to-end mismatch
($\gamma_{\rm\scriptsize jump}<0.004\%$).  Therefore we didn't have to worry
about the periodicity artefact.

The only method that has been proposed so far that strictly implements
$C(\tau)$ rather than $C_p(\tau)$ is given in Ref.~\cite{anneal} and will be
discussed in detail in Sec.~\ref{sec:anneal} below. The method is very accurate
but also rather costly in terms of computer time.  It should be used in cases
of doubt and whenever a suitable sub-sequence cannot be found.

\subsection{Iterative multivariate surrogates}\label{sec:multi1}
\bfig%
\centerline{% 
   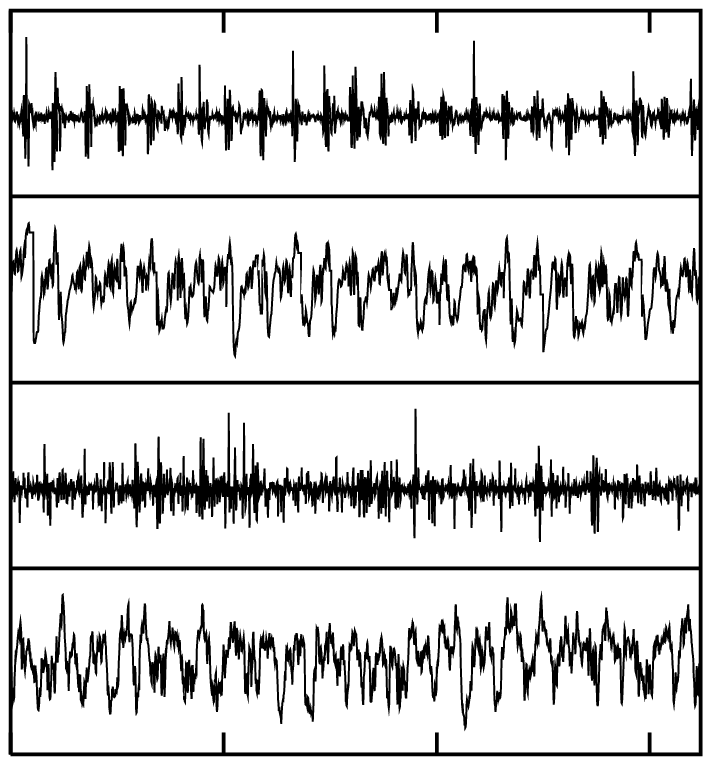% 
}
\caption[]{\label{fig:bdat}
   Simultaneous surrogates for a bi-variate time series. The upper two panels
   show simultaneous recordings of the breath rate and the instantaneous heart
   rate of a human. The lower two panels show surrogate sequences that preserve
   the individual distributions and power spectra as well as the
   cross-correlation function between heart and breath rate. The most prominent
   difference between data and surrogates is the lack of coherence in the
   surrogate breath rate.  }
\end{figure}

\bfig%
\centerline{% 
   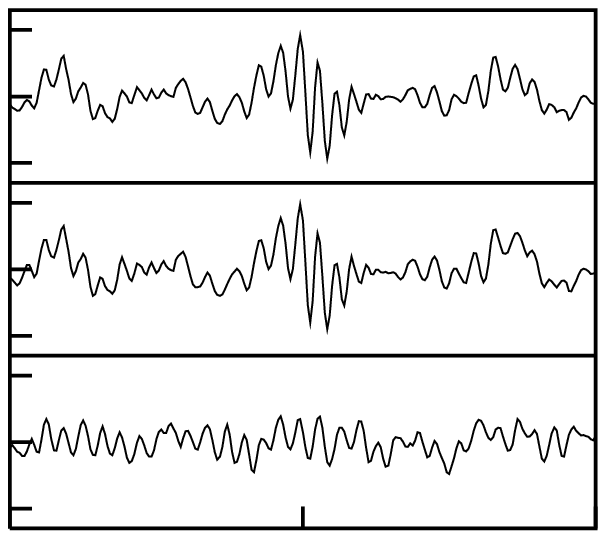% 
}
\caption[]{\label{fig:bx}
   Cross-correlation functions for the bi-variate data shown in
   Fig.~\ref{fig:bdat} (upper panel), and a surrogate that preserves the
   individual spectra and distributions as well as the relative Fourier phases
   (middle). The lower panel shows the same for surrogates prepared for each
   channel individually, that is, without explicitly preserving the
   cross-correlation structure.}
\end{figure}

A natural generalisation of the null hypothesis of a Gaussian linear stochastic
process is that of a multivariate process of the same kind. In this case, the
process is determined by giving the cross-spectrum in
addition to the power spectrum of each of the channels. In Ref.~\cite{multi},
it has been pointed out that phase randomised surrogates are readily produced
by multiplying the Fourier phases of each of the channels by the same set of 
random phases since the cross-spectrum reflects relative phases only.
The authors of Ref.~\cite{multi} did not discuss the possibility to combine 
multivariate phase randomisation with an amplitude adjustment step.
The extension of the iterative refinement scheme introduced in
Sec.~\ref{sec:iterative} to the multivariate case is relatively
straightforward. Since deviations from a Gaussian distribution are very common
and may occur due to a simple invertible rescaling due to the measurement
process, we want to give the algorithm here.

Recall that the iterative scheme consists of two procedures which are applied
in an alternating fashion until convergence to a fixed point is achieved. The
amplitude adjustment procedure by rank ordering (\ref{eq:step2}) is readily
applied to each channel individually. However, the spectral adjustment in the
Fourier domain has to be modified. Let us introduce a second index in order to
denote the $M$ different channels of a multivariate time series $\{s_{n,m},
\quad n=1, \ldots,N, \quad m=1,\ldots,M\}$. The change that has to be applied
to the ``filter'' step, Eq.(\ref{eq:step1}), is that the phases
$\psi_{k,m}$ have to be replaced by phases $\phi_{k,m}$ with the
following properties. (We have dropped the superscript $(i)$ for convenience.)
The replacement should be minimal in the least
squares sense, that is, it should minimise
\be\label{eq:sum}
    h_k = \sum_{m=1}^M \left| e^{i\phi_{k,m}} - e^{i\psi_{k,m}}  \right|^2
\,.\ee
Also, the new phases must implement the same phase differences exhibited by the
corresponding phases $e^{i\rho_{k,m}}=S_{k,m}/ |S_{k,m} |$ of the data:
\be
    e^{i(\phi_{k,m_2} - \phi_{k,m_1})} =
    e^{i(\rho_{k,m_2} - \rho_{k,m_1})}
\,.\ee
The last equation can be fulfilled by setting $\phi_{k,m} = \rho_{k,m} +
\alpha_k$. With this, we have $h_k= \sum_{m=1}^M 2 - 2\cos(\alpha_k - 
\psi_{k,m} + \rho_{k,m})$ which is extremal when
\be
   \tan \alpha_k = {\sum_{m=1}^M \sin (\psi_{k,m} - \rho_{k,m})
         \over \sum_{m=1}^M \cos (\psi_{k,m} - \rho_{k,m})}
\,.\ee
The minimum is selected by taking $\alpha_k$ in the correct quadrant.

As an example, let us generate a surrogate sequence for a simultaneous
recording of the breath rate and the instantaneous heart rate of a
human during sleep. The data is again taken from data set B of the
Santa Fe Institute time series contest~\cite{gold}. The 1944 data
points are an end-point matched sub-sequence of the data used as a
multivariate example in Ref.~\cite{anneal}. In the latter study, which
will be commented on in Sec.~\ref{sec:multi2} below, the breath rate
signal had been considered to be an input and therefore not been
randomised. Here, we will randomise both channels under the condition
that their individual spectra as well as their cross-correlation
function are preserved as well as possible while matching the
individual distributions exactly. The iterative scheme introduced
above took 188 iterations to converge to a fixed point.  The data and
a bi-variate surrogate is shown in Fig.~\ref{fig:bdat}.  In
Fig.~\ref{fig:bx}, the cross-correlation functions of the data and one
surrogate are plotted. Also, for comparison, the same for two individual
surrogates of the two channels. The most striking
difference between data and surrogates is that the coherence of the
breath rate is lost. Thus, it is indeed reasonable to exclude the
nonlinear structure in the breath dynamics from a further analysis of
the heart rate by taking the breath rate as a given input signal.
Such an analysis is however beyond the scope of the method discussed
in this section. First of all, specifying the full cross-correlation
function to a fixed signal plus the autocorrelation function
over-specifies the problem and there is no room for randomisation. In
Sec.~\ref{sec:multi2} below, we will therefore revisit this
problem. With the general constrained randomisation scheme to be
introduced below, it will be possible to specify a limited number of
lags of the auto- and cross-correlation functions.

\section{General constrained randomisation}\label{sec:anneal}
Randomisation schemes based on the Fourier amplitudes of the data are
appropriate in many cases. However, there remain some flaws, the strongest
being the severely restricted class of testable null hypotheses.  The
periodogram estimator of the power spectrum is about the only interesting
observable that allows for the solution of the inverse problem of generating
random sequences under the condition of its given value.

In the general approach of Ref.~\cite{anneal}, constraints ({\sl e.g.}
autocorrelations) on the surrogate data are implemented by a cost function
which has a global minimum when the constraints are fulfilled. This general
framework is much more flexible than the Fourier based methods. We will
therefor discuss it in some detail.

\subsection{Null hypotheses, constraints, and cost functions}
As we have discussed previously, we will often have to specify a null
hypothesis in terms of a complete set of observable properties of the data.
Only in specific cases (e.g. the two point autocorrelation function), there is
a one-to-one correspondence to a class of models (here the ARMA process).  In
any case, if $\{\overline{s}_n\}$ denotes a surrogate time series, the
constraints will most often be of (or can be brought into) the form
\be\label{eq:F}
   F_i(\{\overline{s}_n\})=0,\quad i=1,\ldots, I
\,.\ee
Such constraints can always be turned into a cost function
\be\label{eq:E}
   E(\{\overline{s}_n\})=\left( 
              \sum_{i=1}^{I} |w_i  F_i(\{\overline{s}_n\}) |^q \right)^{1/q}
\,.\ee
The fact that $E(\{\overline{s}_n\})$ has a global minimum when the constraints
are fulfilled is unaffected by the choice of the weights $w_i\ne 0$ and the
order $q$ of the average.  The least squares or $L^2$ average is obtained at
$q=2$, $L^1$ at $q=1$ and the maximum distance when $q\to\infty$. Geometric
averaging is also possible (and can be formally obtained by taking the limit
$q\to 0$ in a proper way).  We have experimented with different choices of $q$
but we haven't found a choice that is uniformly superior to others. It seems
plausible to give either uniform weights or to enhance those constraints which
are particularly difficult to fulfil. Again, conclusive empirical results are
still lacking.

Consider as an example the constraint that the sample autocorrelation function
of the surrogate $\overline{C}(\tau)=\av{\overline{s}_n \overline{s}_{n-\tau}}$
(data rescaled to zero mean and unit variance) are the same as those of the
data, $C(\tau)=\av{s_n s_{n-\tau}}$. This is done by specifying zero
discrepancy as a constraint $F_\tau(\{\overline{s}_n\}) =
\overline{C}(\tau)-C(\tau), \quad \tau=1,\ldots,\tau_{\rm\scriptsize max}$. If
the correlations decay fast, $\tau_{\rm\scriptsize max}$ can be restricted,
otherwise $\tau_{\rm\scriptsize max}=N-1$ (the largest available lag).  Thus, a
possible cost function could read
\be\label{eq:cost}
   E= {\rm max}_{\tau=0}^{\tau_{\rm\scriptsize max}} 
      \left|\overline{C}(\tau)-C(\tau) \right|
\,.\ee
Other choices of $q$ and the weights are of course also possible.

In all the cases considered in this paper, one constraint will be that the
surrogates take on the same values as the data but in different time order.
This ensures that data and surrogates can equally likely be drawn from the same
(unknown) single time probability distribution. This particular constraint is
not included in the cost function but identically fulfilled by considering only
permutations without replacement of the data for minimisation.

By introducing a cost function, we have turned a difficult nonlinear, high
dimensional root finding problem (\ref{eq:F}) into a minimisation problem
(\ref{eq:E}). This leads to extremely many false minima whence such a strategy
is discouraged for general root finding problems~\cite{Press92}. Here, the
situation is somewhat different since we need to solve Eq.(\ref{eq:F}) only
over the set of all permutations of $\{s_n\}$. Although this set is big, it is
still discrete and powerful combinatorial minimisation algorithms are
available that can deal with false minima very well. We choose to minimise
$E(\{\overline{s}_n\})$ among all permutations $\{\overline{s}_n\}$ of the
original time series $\{s_n\}$ using the method of {\em simulated annealing}.
Configurations are updated by exchanging pairs in $\{\overline{s}_n\}$. The
annealing scheme will decide which changes to accept and which to reject. With
an appropriate cooling scheme, the annealing procedure can reach any desired
accuracy. Apart from simulated annealing, genetic algorithms~\cite{genetic}
have become very popular for this kind of problems and there is no reason why
they couldn't be used for the present purpose as well.

\subsection{Computational issues of simulated annealing}
Simulated annealing is a very powerful method of combinatorial minimisation in
the presence of many false minima. Simulated annealing has a rich literature,
classical references are Metropolis et al.~\cite{metro} and
Kirkpatrick~\cite{kirk}, more recent material can be found for example in
Vidal~\cite{annealbook}. Despite its many successful applications, using
simulated annealing efficiently is still a bit of an art. We will here discuss
some issues we have found worth dealing with in our particular minimisation
problem. Since the detailed behaviour will be different for each cost function,
we can only give some general guidelines.

\bfig%
\centerline{% 
   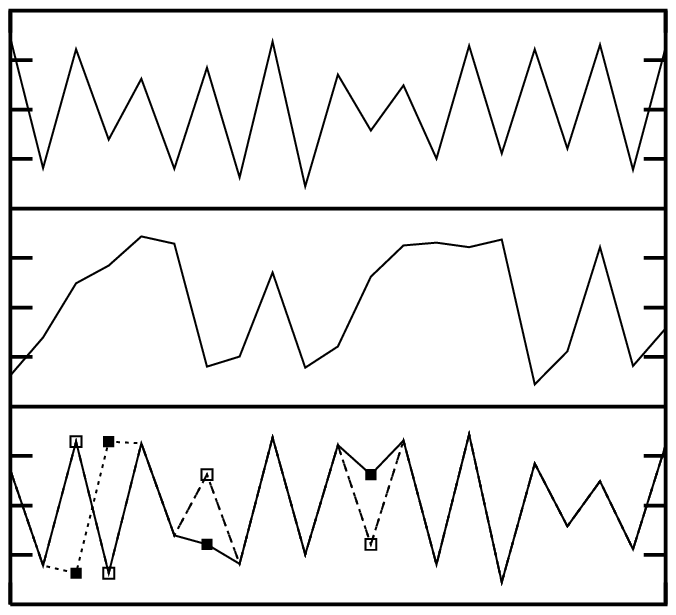% 
}
\caption[]{\label{fig:update}
   Building up correlations by pairwise permutation. Suppose we want to
   generate the strong anti-correlation present in the data (upper trace) by
   minimising $E=|\overline{C}(1)-C(1)|$. The annealing started with a random
   permutation (middle trace, $E=1.129$).  At a given intermediate state (lower
   trace, $E=0.256$), exchanging the points $a$ and $b$ increases the
   cost to $E=0.2744$ while exchanging $c$ and $d$ creates negative correlation
   and reduces the cost to $E=0.002$.}
\end{figure}
 
The main idea behind simulated annealing is to interpret the cost function $E$
as an energy in a thermodynamic system. Minimising the cost function is then
equivalent to finding the ground state of a system. A glassy solid can be
brought close to the energetically optimal state by first heating it and
subsequently cooling it. This procedure is called ``annealing'', hence the name
of the method.  If we want to simulate the thermodynamics of this tempering
procedure on a computer, we notice that in thermodynamic equilibrium at some
finite temperature $T$, system configurations should be visited with a
probability according to the Boltzmann distribution $e^{-E/T}$ of the canonical
ensemble. In Monte Carlo simulations, this is achieved by accepting changes of
the configuration with a probability $p=1$ if the energy is decreased $(\Delta
E<0)$ and $p=e^{-\Delta E/T}$ if the energy is increased, $(\Delta E\ge
0)$. This selection rule is often referred to as the {\em Metropolis step}. In
a minimisation problem, the temperature is the parameter in the Boltzmann
distribution that sets its width. In particular, it determines the probability
to go ``up hill'', which is important if we need to get out of false minima.

In order to anneal the system to the ground state of minimal ``energy'', that
is, the minimum of the cost function, we want to first ``melt'' the system at a
high temperature $T$, and then decrease $T$ slowly, allowing the system to be
close to thermodynamic equilibrium at each stage. If the changes to the
configuration we allow to be made connect all possible states of the system,
the updating algorithm is called {\em ergodic}.  Although some general rigorous
convergence results are available, in practical applications of simulated
annealing some problem-specific choices have to be made. In particular, apart
from the constraints and the cost function, one has to specify a method of
updating the configurations and a schedule for lowering the temperature. In the
following, we will discuss each of these issues.

Concerning the choice of cost function, we have already mentioned that there is
a large degeneracy in that many cost functions have an absolute minimum
whenever a given set of constraints if fulfilled. The convergence properties
can depend dramatically on the choice of cost function. Unfortunately, this
dependence seems to be so complicated that it is impossible even to discuss the
main behaviour in some generality. In particular, the weights $w_i$ in
Eq.(\ref{eq:E}) are sometimes difficult to choose. Heuristically, we would like
to reflect changes in the $I$ different constraints about equally, provided the
constraints are independent. Since their scale is not at all set by
Eq.(\ref{eq:F}), we can use the $w_i$ for this purpose. Whenever we have some
information about which kind of deviation would be particularly problematic
with a given test statistic, we can give it a stronger weight. Often, the
shortest lags of the autocorrelation function are of crucial importance, whence
we tend to weight autocorrelations by $1/\tau$ when they occur in sums. Also,
the $C(\tau)$ with larger $\tau$ are increasingly ill-determined due to the
fewer data points entering the sums. As an extreme example, $C(N-1)=s_1s_{N-1}$
shows huge fluctuations due to the lack of self-averaging. Finally, there are
many more $C(\tau)$ with larger $\tau$ than at the crucial short lags.

A way to efficiently reach all permutations by small individual changes is by
exchanging randomly chosen (not necessarily close-by) pairs. How the
interchange of two points can affect the current cost is illustrated
schematically in Fig.~\ref{fig:update}.  Optimising the code that computes and
updates the cost function is essential since we need its current value at each
annealing step --- which are expected to be many.  Very often, an exchange of
two points is reflected in a rather simple update of the cost function. For
example, computing $C(\tau)$ for a single lag $\tau$ involves $O(N)$
multiplications. Updating $C(\tau)$ upon the exchange of two points $i<j$ only
requires the replacement of the terms $s_is_{i-\tau}$, $s_{i+\tau}s_{i}$,
$s_js_{j-\tau}$, and $s_{j+\tau}s_j$ in the sum. Note that cheap updates are a
source of potential mistakes (e.g. avoid subtracting terms twice in the case
that $i=j-\tau$) but also of roundoff errors. To ensure that the assumed cost
is always equal to the actual cost, code carefully and monitor roundoff by
computing a fresh cost function occasionally.

Further speed-up can be achieved in two ways. Often, not all the terms in a
cost function have to be added up until it is clear that the resulting change
goes up hill by an amount that will lead to a rejection of the exchange.
Also, pairs to be exchanged can be selected closer in magnitude at low
temperatures because large changes are very likely to increase the cost.

Many cooling schemes have been discussed in the
literature~\cite{annealbook}. We use an exponential scheme in our work.  We
will give details on the -- admittedly largely ad hoc --- choices that have
been made in the TISEAN implementation in Appendix~\ref{app:A}. We found it
convenient to have a scheme available that automatically adjusts parameters
until a given accuracy is reached. This can be done by cooling at a certain
rate until we are stuck (no more accepted changes). If the cost is not low
enough yet, we melt the system again and cool at a slower rate.

\subsection{Example: avoiding periodicity artefacts}
\bfig%
\centerline{% 
   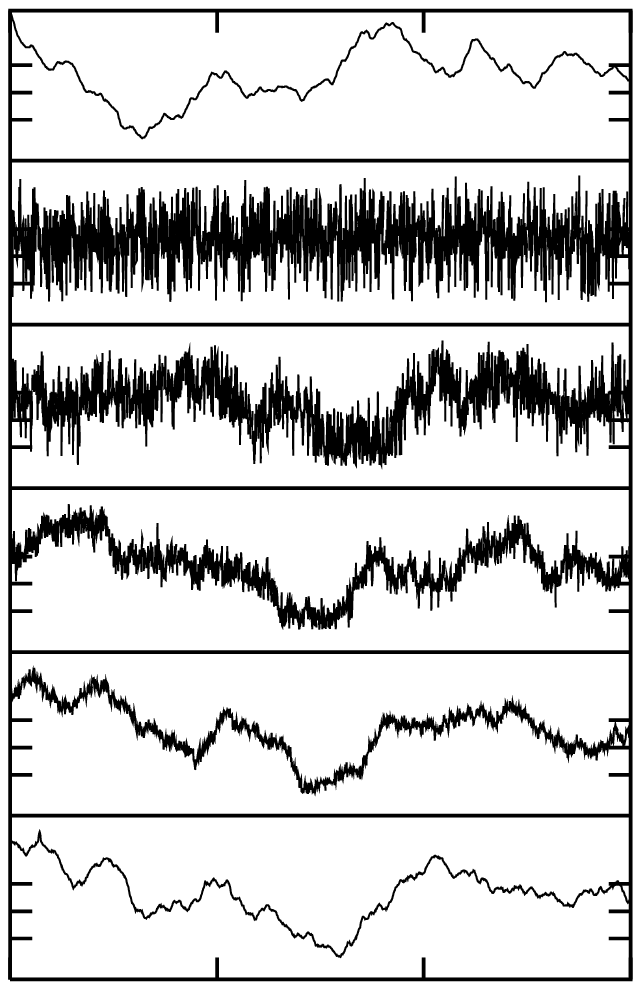% 
}
\caption[]{\label{fig:endanneal}
   Progressive stages of the simulated annealing scheme. The data used in
   Fig.~\ref{fig:end} is used to generate an annealed surrogate that minimises
   $E= \mbox{max}_{\tau=0}^{100} \,|\overline{C}(\tau)-C(\tau)|$ over all
   permutations of the data. From top to bottom, the values for $E$ are: 0
   (original data), 1.01 (random scramble), 0.51, 0.12, 0.015, and 0.00013.}
\end{figure}

Let us illustrate the use of the annealing method in the case of the standard
null hypothesis of a rescaled linear process. We will show how the periodicity
artefact discussed in Sec.~\ref{sec:period} can be avoided by using a more
suitable cost function. We prepare a surrogate for the data shown in
Fig.~\ref{fig:end} (almost unstable AR(2) process) without truncating its
length. We minimise the cost function given by Eq.(\ref{eq:cost}), involving
all lags up to $\tau_{\rm\scriptsize max}=100$. Also, we excluded the first and
last points from permutations as a cheap way of imposing the long range
correlation. In Fig.~\ref{fig:endanneal} we show progressive stages of the
annealing procedure, starting from a random scramble. The temperature $T$ is
decreased by 0.1\%  after either $10^6$ permutations have been tried or $10^4$
have been successful.  The final surrogate neither has spuriously matching ends
nor the additional high frequency components we saw in Fig.~\ref{fig:end}. The
price we had to pay was that the generation of one single surrogate took 6~h of
CPU time on a Pentium~II PC at 350~MHz.  If we had taken care of the long range
correlation by leaving the end points loose but taking $\tau_{\rm\scriptsize
max}=N-1$, convergence would have been prohibitively slow.  Note that for a
proper test, we would need at least 19 surrogates. We should stress that this
example with its very slow decay of correlations is particularly nasty --- but
still feasible. Obviously, sacrificing 10\%  of the points to get rid of the
end point mismatch is preferable here to spending several days of CPU time
on the annealing scheme. In other cases, however, we may not have such a
choice. 

\subsection{Combinatorial minimisation and accuracy}
In principle, simulated annealing is able to reach arbitrary accuracy at the
expense of computer time. We should, however, remark on a few points.  Unlike
other minimisation problems, we are not really interested in the solutions that
put $E=0$ exactly. Most likely, these are the data set itself and a few simple
transformations of it that preserve the cost function (e.g. a time reversed
copy). On the other hand, combinatorics makes it very unlikely that we ever
reach one of these few of the $N!$ permutations, unless $N$ is really small or
the constraints grossly over-specify the problem.  This can be the case, for
example, if we include {\em all} possible lags of the autocorrelation function,
which gives as many (nonlinear) equations as unknowns, $I=N$. These may close
for small $N$ in the space of permutations.  In such extreme situations, it is
possible to include extra cost terms penalising closeness to one of the trivial
transformations of the data.  Let us note that if the surrogates are ``too
similar'' to the data, this does not in itself affect the validity of the
test. Only the discrimination power may be severely reduced.

Now, if we don't want to reach $E=0$, how can we be sure that there are enough
independent realisations with $E\approx 0$? The theoretical answer depends on
the form of the constraints in a complicated way and cannot be given in
general. We can, however, offer a heuristic argument that the number of
configurations with $E$ smaller than some $\Delta E$ grows fast for large $N$.
Suppose that for large $N$ the probability distribution of $E$ converges to an
asymptotic form $p(E)$. Assume further that $\tilde{p}(\Delta
E)=\mbox{Prob}(\,E<\Delta E\,)=\int_0^{\Delta E} p(E)dE$ is nonzero but maybe
very small. This is evidently true for autocorrelations, for example. While
thus the probability to find $E<\Delta E$ in a random draw from the
distribution of the data may be extremely small, say $\tilde{p}(\Delta
E)=10^{-45}$ at 10 sigmas from the mean energy, the total number of
permutations, figuring as the number of draws, grows as $N!\approx
(N/e)^N\sqrt{2\pi N}$, that is, much faster than exponentially. Thus, we expect
the number of permutations with $E<\Delta E$ to be $\propto \tilde{p}(\Delta E)
N!$.  For example, $10^{-45}\times1000!\approx 10^{2522}$.

In any case, we can always monitor the convergence of the cost function to
avoid spurious results due to residual inaccuracy in the surrogates.  As we
will discuss below, it can also be a good idea to test the surrogates with a
linear test statistic before performing the actual nonlinearity test.

\subsection{The curse of accuracy}\label{sec:accuracy}
Strictly speaking, the concept of constrained realisations requires the
constraints to be fulfilled {\em exactly}, a practical impossibility. Most of
the research efforts reported in this article have their origin in the attempt
to increase the accuracy with which the constraints are implemented, that is,
to minimise the bias resulting from any remaining discrepancy. Since most
measures of nonlinearity are also sensitive to linear correlations, a side
effect of the reduced bias is a reduced variance of such
estimators. Paradoxically, thus the enhanced accuracy may result in false
rejections of the null hypothesis on the ground of tiny differences in some
nonlinear characteristics. This important point has been recently put forth by
Kugiumtzis~\cite{dimitris}.

Consider the highly correlated autoregressive process $x_n = 0.99x_{n-1} -
0.8x_{n-2} + 0.65x_{n-3} + \eta_n$, measured by the function
$s_n=s(x_n)=x_n|x_n|$ and then normalised to zero mean and unit variance.  The
strong correlation together with the rather strong static nonlinearity makes
this a very difficult data set for the generation of
surrogates. Figure~\ref{fig:accuracy} shows the bias and variance for a {\em
linear} statistic, the unit lag autocorrelation $C_p(1)$, Eq.(\ref{eq:cp}), as
compared to its goal value given by the data.  The left part of
Fig.~\ref{fig:accuracy} shows $C_p(1)$ versus the iteration count $i$ for 200
iterative surrogates, $i=1$ roughly corresponding to AAFT surrogates.  Although
the mean accuracy increases dramatically compared to the first iteration
stages, the data consistently remains outside a 2$\sigma$ error bound.  Since
nonlinear parameters will also pick up linear correlations, we have to expect
spurious results in such a case. In the right part, annealed surrogates are
generated with a cost function $E=\max_{\tau=1}^{200}
|\overline{C}_p(\tau)-C_p(\tau)|/\tau$. The bias and variance of $C_p(1)$ are
plotted versus the cost~$E$. Since the cost function involves $C_p(1)$, it is
not surprising that we see good convergence of the bias. It is also noteworthy
that the variance is in any event large enough to exclude spurious results due
to remaining discrepancy in the linear correlations.

\bfig%
\centerline{% 
   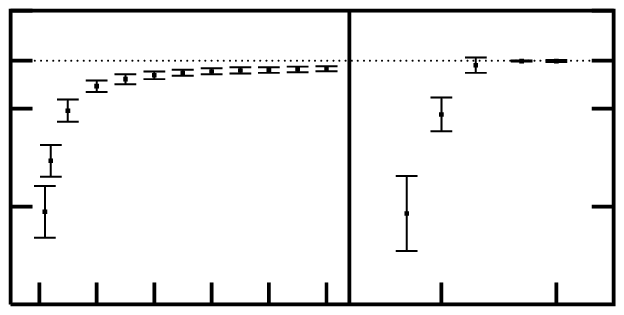% 
}
\caption[]{\label{fig:accuracy}
   Bias and variance of unit lag autocorrelation $C_p(1)$ for ensembles of
   surrogates.  Left part: $C_p(1)$ plotted versus the iteration count $i$ for
   200 iterative surrogates. The AAFT method gives accuracies comparable to the
   value obtained for $i=1$.  Right part: $C_p(1)$ plotted versus the goal
   value of the cost function for 20 annealed surrogates.  The horizontal line
   indicates the sample value for the data sequence. See text for discussion.}
\end{figure}

Kugiumtzis~\cite{dimitris} suggests to test the validity of the surrogate
sample by performing a test using a linear statistic for normalisation. For the
data shown in Fig.~\ref{fig:accuracy}, this would have detected the lack of
convergence of the iterative surrogates.  Currently, this seems to be the only
way around the problem and we thus recommend to follow his suggestion. With the
much more accurate annealed surrogates, we haven't so far seen examples of
dangerous remaining inaccuracy, but we cannot exclude their possibility. If
such a case occurs, it may be possible to generate unbiased ensembles of
surrogates by specifying a cost function that explicitly minimises the bias.
This would involve the whole collection of $M$ surrogates at the same time,
including extra terms like
\be
    E_{\rm\scriptsize ensemble} = \sum_{\tau=0}^{\tau_{\rm\scriptsize max}} 
       \left(\sum_{m=1}^M \overline{C}_m(\tau)-C(\tau) \right)^2
\,.\ee
Here, $\overline{C}_m(\tau)$ denotes the autocorrelation function of the
$m$--th surrogate.  In any event, this will be a very cumbersome procedure, in
terms of implementation and in terms of execution speed and it is questionable
if it is worth the effort.

\section{Various Examples}
In this section, we want to give a number of applications of the constrained
randomisation approach. If the constraints consist only of the Fourier
amplitudes and the single time probability distribution, the iteratively
refined, amplitude adjusted surrogates~\cite{surrowe} discussed in
Sec.~\ref{sec:iterative} are usually sufficient if the end point artefact can
be controlled and convergence is satisfactory. Even the slightest extension of
these constraints makes it impossible to solve the inverse problem directly and
we have to follow the more general combinatorial approach discussed in the
previous section.  The following examples are meant to illustrate how this can
be carried out in practice.

\subsection{Including non-stationarity}\label{sec:including}
\bfig%
   \centerline{% 
   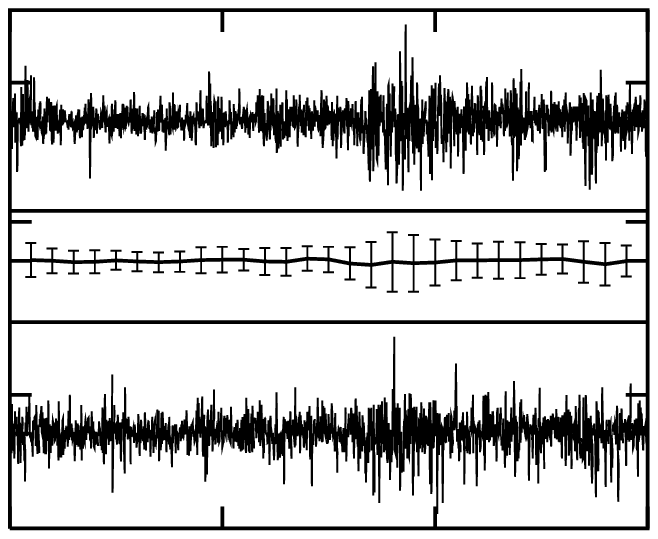% 
}
   \caption[]{Non-stationary financial time series (BUND Future returns, top)
      and a surrogate (bottom) preserving the non-stationary structure
      quantified by running window estimates of the local mean and variance
      (middle).\label{fig:bund}}
\end{figure}

Constrained randomisation using combinatorial minimisation is a very flexible
method since in principle arbitrary constraints can be realised. Although it
is seldom possible to specify a formal null hypothesis for more general
constraints, it can be quite useful to be able to incorporate into the
surrogates any feature of the data that is understood already or that is
uninteresting. Non-stationarity has been excluded so far by requiring the
equations defining the null hypothesis to remain constant in time. This has a
two-fold consequence. First, and most importantly, we must keep in mind that
the test will have discrimination power against non-stationary signals as a
valid alternative to the null hypothesis. Thus a rejection can be due to
nonlinearity or non-stationarity equally well.

Second, if we do want to include non-stationarity in the null hypothesis we
have to do so explicitly. Let us illustrate how this can be done with an
example from finance. The time series consists of 1500 daily returns (until the
end of 1996) of the {\em BUND Future}, a derived German financial instrument.
The data were kindly provided by Thomas Sch\"urmann, WGZ-Bank D\"usseldorf. As
can be seen in the upper panel of Fig.~\ref{fig:bund}, the sequence is
non-stationary in the sense that the local variance and to a lesser extent also
the local mean undergo changes on a time scale that is long compared to the
fluctuations of the series itself. This property is known in the statistical
literature as {\em heteroscedasticity} and modelled by the so-called
GARCH~\cite{garch} and related models. Here, we want to avoid the construction
of an explicit model from the data but rather ask the question if the data is
compatible with the null hypothesis of a correlated linear stochastic process
with time dependent local mean and variance. We can answer this question in a
statistical sense by creating surrogate time series that show the same linear
correlations and the same time dependence of the running mean and running
variance as the data and comparing a nonlinear statistic between data and
surrogates. The lower panel in Fig.~\ref{fig:bund} shows a surrogate time
series generated using the annealing method.  The cost function was set up to
match the autocorrelation function up to five days and the moving mean and
variance in sliding windows of 100 days duration.  In Fig.~\ref{fig:bund} the
running mean and variance are shown as points and error bars, respectively, in
the middle trace. The deviation of these between data and surrogate has been
minimised to such a degree that it can no longer be resolved.  A comparison of
the time-asymmetry statistic Eq.(\ref{eq:skew}) for the data and 19 surrogates
did not reveal any discrepancy, and the null hypothesis could not be rejected.

\subsection{Multivariate data}\label{sec:multi2} 
In Ref.~\cite{anneal}, the flexibility of the approach was illustrated by a
simultaneous recording of the breath rate and the instantaneous heart rate of a
human subject during sleep. The interesting question was, how much of the
structure in the heart rate data can be explained by linear dependence on the
breath rate. In order to answer this question, surrogates were made that had
the same autocorrelation structure but also the same cross-correlation with
respect to the fixed input signal, the breath rate. While the linear
cross-correlation with the breath rate explained the coherent structure of the
heart rate, other features, in particular its asymmetry under time reversal,
remained unexplained. Possible explanations include artefacts due to the
peculiar way of deriving heart rate from inter-beat intervals, nonlinear
coupling to the breath activity, nonlinearity in the cardiac system, and
others. 

Within the general framework, multivariate data can be treated very much the 
same way as scalar time series. In the above example, we chose to use one of
the channels as a reference signal which was not randomised. The rationale
behind this was that we were not looking for nonlinear structure in the breath
rate itself and thus we didn't want to destroy any such structure in the
surrogates. In other cases, we can decide either to keep or to destroy
cross-correlations between channels. The former can be achieved by applying the
same permutations to all channels. Due to the limited experience we have so far
and the multitude of possible cases, multivariate problems have not been
included in the TISEAN implementation yet.

\subsection{Uneven sampling}\label{sec:uneven}
Let us show how the constrained randomisation method can be used to test for
nonlinearity in time series taken at time intervals of different length.
Unevenly sampled data are quite common, examples include drill core
data, astronomical observations or stock price notations. Most observables and
algorithms cannot easily be generalised to this case which is particularly true
for nonlinear time series methods. (See~\cite{XParzen83} for material on
irregularly sampled time series.) Interpolating the data to equally spaced
sampling times is not recommendable for a test for nonlinearity since one could
not {\sl a posteriori} distinguish between genuine structure and nonlinearity
introduced spuriously by the interpolation process. Note that also zero padding
is a nonlinear operation in the sense that stretches of zeroes are unlikely to
be produced by any linear stochastic process.

For data that is evenly sampled except for a moderate number of gaps, surrogate
sequences can be produced relatively straightforwardly by assuming the value
zero during the gaps and minimising a standard cost function like
Eq.(\ref{eq:cost}) while excluding the gaps from the permutations tried. The
error made in estimating correlations would then be identical for the data and
surrogates and could not affect the validity of the test. Of course, one would
have to modify the nonlinearity measure to avoid the gaps. For data sampled at
incommensurate times, such a strategy can no longer be adopted. We then need
different means to specify the linear correlation structure.

Two different approaches are viable, one residing in the spectral domain and
one in the time domain.  Consider a time series sampled at times $\{t_n\}$ that
need not be equally spaced. The power spectrum can then be estimated by the
Lomb periodogram, as discussed for example in Ref.~\cite{Press92}.
For time series sampled at constant time intervals, the Lomb periodogram yields
the standard squared Fourier transformation.  Except for this particular case,
it does not have any inverse transformation, which makes it impossible to use
the standard surrogate data algorithms mentioned in Sec.~\ref{sec:fourier}.  In
Ref.~\cite{lomb}, we used the Lomb periodogram of the data as a constraint for
the creation of surrogates.  Unfortunately, imposing a given Lomb periodogram
is very time consuming because at each annealing step, the $O(N)$ spectral
estimator has to be computed at $O(N_f)$ frequencies with $N_f \propto N$.
Press et al.~\cite{Press92} give an approximation algorithm that uses the fast
Fourier transform to compute the Lomb periodogram in $O(N\log N)$ time rather
than $O(N^2)$. The resulting code is still quite slow.

As a more efficient alternative to the commonly used but computationally costly
Lomb periodogram, let us suggest to use binned autocorrelations. They are
defined as follows. For a continuous signal $s(t)$ (take $\av{s}=0$,
$\av{s^2}=1$ for simplicity of notation here), the autocorrelation function is
$C(\tau)=\av{s(t)s(t-\tau)}=(1/T)\int_{0}^T\!dt'\, s(t')s(t'-\tau)$. It can be
binned to a bin size $\Delta$, giving $C_{\Delta}(\tau)=(1/\Delta)
\int_{\tau-\Delta}^{\tau}d\tau'\, C(\tau')$. We now have to approximate all
integrals using the available values of $s(t_n)$. In general, we estimate
\be
    \int_a^b \!dt\; f(t) \approx (b-a)
        { \sum_{{\cal B}_n(a,b)}\; f(t_n) \over 
            |{\cal B}_n(a,b)| }
\,.\ee
Here, ${\cal B}_n(a,b)=\{n:\; a < t_n\le b\}$ denotes the bin ranging from $a$
to $b$ and $|{\cal B}(a,b)|$ the number of its elements. We could improve this
estimate by some interpolation of $f(\cdot)$, as it is customary with numerical
integration but the accuracy of the estimate is not the central issue here.
For the binned autocorrelation, this approximation simply gives
\be\label{eq:cdelta}
    C_{\Delta}(\tau)\approx
        { \sum_{{\cal B}_{ij}(\tau-\Delta,\tau)} \;s(t_i)s(t_j) \over
           |{\cal B}_{ij}(\tau-\Delta,\tau)| }
\,.\ee
Here, ${\cal B}_{ij}(a,b)=\{(i,j):\; a < t_i-t_j\le b\}$.
Of course, empty bins lead to undefined autocorrelations.  If we have evenly
sampled data and unit bins, $t_i-t_{i-1}=\Delta, \quad i=2,\ldots N$, then the
binned autocorrelations coincide with ordinary autocorrelations at
$\tau=i\Delta, \quad i=0,\ldots,N-1$.

Once we are able to specify the linear properties of a time series, we can also
define a cost function as usual and generate surrogates that realise the binned
autocorrelations of the data. A delicate point however is the choice of bin
size. If we take it too small, we get bins that are almost empty. Within the
space of permutations, there may be only a few ways then to generate precisely
that value of $\overline{C}_{\Delta}(\tau)$, in other words, we over-specify
the problem. If we take the bin size too large, we might not capture important
structure in the autocorrelation function.
 
As an application, let us construct randomised versions of part of an ice core
data set, taken from the Greenland Ice Sheet Project Two (GISP2)~\cite{gisp2}.
An extensive data base resulting from the analysis of physical and chemical
properties of Greenland ice up to a depth of 3028.8~m has been published by the
National Snow and Ice Data Center together with the World Data Center-A for
Palaeoclimatology, National Geophysical Data Center, Boulder,
Colorado~\cite{CDROM}. A long ice core is usually cut into equidistant slices
and initially, all measurements are made versus depth. Considerable expertise
then goes into the dating of each slice~\cite{dating}. Since the density of the
ice, as well as the annual total deposition, changes with time, the final time
series data are necessarily unevenly sampled. Furthermore, often a few values
are missing from the record.  We will study a subset of the data ranging back
10000~years in time, corresponding to a depth of 1564~m, and continuing until
2000~years before present. Figure~\ref{fig:icetime} shows the sampling rate
versus time for the particular ice core considered.

\bfig%
   \centerline{% 
   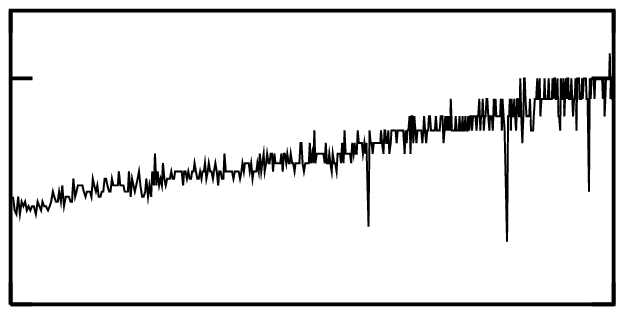% 
}
   \caption[]{\label{fig:icetime} 
      Sampling rate versus time for an ice core time series.}
\end{figure}

We use the $\delta^{18}$O time series which indicates the deviation of the
$\alpha=$ $^{18}{\rm O}/^{16}{\rm O}$ ratio from its standard value $\alpha_0$:
$\delta^{18}\mbox{O}=0.103 (\alpha-\alpha_0)/\alpha_0$. Since the ratio of the
condensation rates of the two isotopes depends on temperature, the isotope
ratio can be used to derive a temperature time series. The upper trace in
Fig.~\ref{fig:ice} shows the recording from 10000~years to 2000~years before
present, comprising 538 data points.

In order to generate surrogates with the same linear properties, we estimate
autocorrelations up to a lag of $\tau=1000$~years by binning to a resolution of
5~y. A typical surrogate is shown as the lower trace in Fig.~\ref{fig:ice}.  We
have not been able to detect any nonlinear structure by comparing this
recording with 19 surrogates, neither using time asymmetry nor prediction
errors. It should be admitted, however, that we haven't attempted to provide
nonlinearity measures optimised for the unevenly sampled case. For that
purpose, also some interpolation is permissible since it is then part of the
nonlinear statistic. Of course, in terms of geophysics, we are asking a very
simplistic question here. We wouldn't really expect strong nonlinear signatures
or even chaotic dynamics in such a single probe of the global climate.  All the
interesting information --- and expected nonlinearity --- lies in the
interrelation between various measurements and the assessment of long term
trends we have deliberately excluded by selecting a subset of the data.

\bfig%
\centerline{% 
   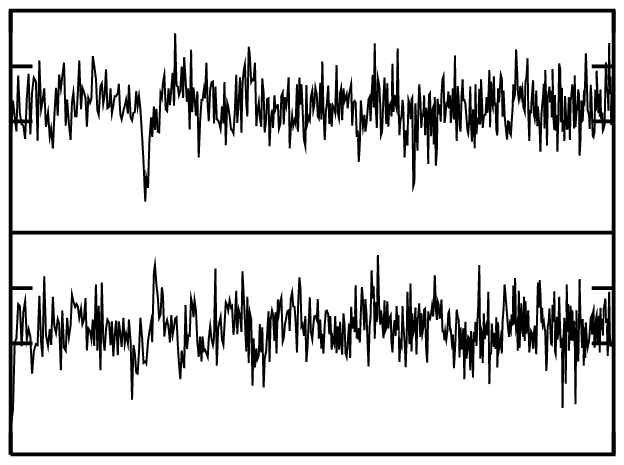% 
}
\caption[]{\label{fig:ice} 
   Oxygen isotope ratio time series derived from an ice core (upper trace) and
   a corresponding surrogate (lower trace) that has the same binned
   autocorrelations up to a lag of 1000~years at a resolution of 5~years.}
\end{figure}

\subsection{Spike trains} 

\bfig%
\centerline{% 
   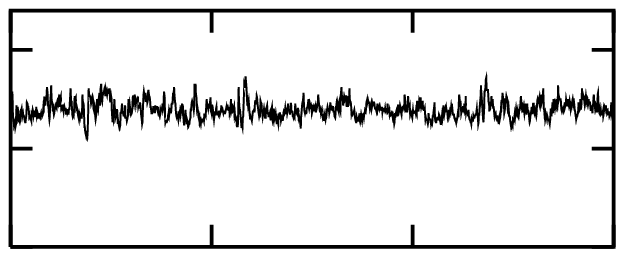% 
}
\caption[]{\label{fig:rrseries}
   Heart rate fluctuations seen by plotting the time interval between
   consecutive heart beats (R waves) versus the beat number. Note that the
   spread of values is rather small due to the near-periodicity of the heart
   beat. }
\end{figure}

\bfig%
\centerline{% 
   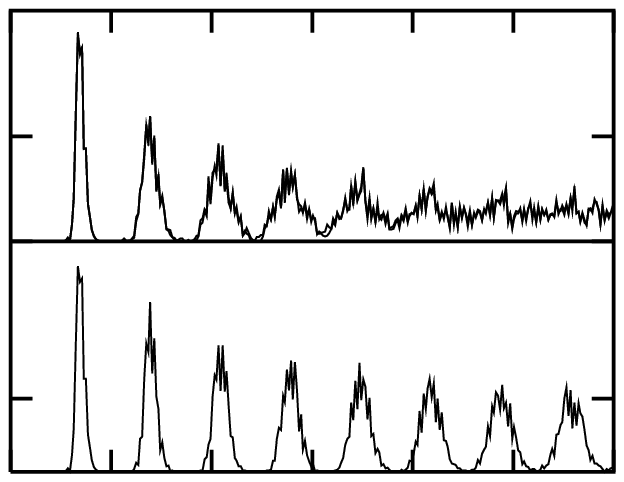% 
}
\caption[]{\label{fig:rr}
   Binned autocorrelation function of an RR interval time series. Upper panel:
   $C(\tau)$ and $\overline{C}(\tau)$ are practically indistinguishable. Lower:
   Autocorrelation for a random scramble of the data. Note that most of the
   periodicity is given by the fact that the duration of each beat had rather
   little variation during this recording.}
\end{figure}

A spike train is a sequence of $N$ events (for example neuronal spikes, or
heart beats) occurring at times $\{t_n\}$. Variations in the events beyond their
timing are ignored.  Let us first note that this very common kind of data is
fundamentally different from the case of unevenly {\em sampled} time series
studied in the last section in that the sampling instances $\{t_n\}$ are not
independent of the measured process. In fact, between these instances, the
value of $s(t)$ is undefined and the $\{t_n\}$ contain all the information
there is.

Very often, the discrete sequence if inter-event intervals $x_n=t_n-t_{n-1}$ is
treated as if it were an ordinary time series. We must keep in mind, however,
that the index $n$ is not proportional to time any more. It depends on the
nature of the process if it is more reasonable to look for correlations in time
or in event number.  Since in the latter case we can use the standard machinery
of regularly sampled time series, let us concentrate on the more difficult real
time correlations.

In particular the literature on {\em heart rate variability} (HRV) contains
interesting material on the question of spectral estimation and linear modeling
of spike trains, here usually inter-beat (RR) interval series, see
e.g. Ref.~\cite{spikespec}. For the heart beat interval sequence shown in
Fig.~\ref{fig:rrseries}, spectral analysis of $x_n$ versus $n$ may reveal
interesting structure, but even the mean periodicity of the heart beat would be
lost and serious aliasing effects would have to be faced.  A very convenient
and powerful approach that uses the real time $t$ rather than the event number
$n$ is to write a spike train as a sum of Dirac delta functions located at the
spike instances:
\be\label{eq:delta}
   s(t)=\sum_{n=1}^N \delta(t-t_n)
\,.\ee
With $\int\!dt\; s(t)e^{i\omega t}=\sum_{n=1}^N e^{-i\omega t_n}$, the
periodogram spectral estimator is then simply obtained by squaring the
(continuous) Fourier transform of $s(t)$:
\be\label{eq:spower}
   P(\omega)={1\over 2\pi} \left|\sum_{n=1}^N e^{-i\omega t_n}\right|^2
\,.\ee
Other spectral estimators can be derived by smoothing $P(\omega)$ or by data
windowing. It is possible to generate surrogate spike trains that realise the
spectral estimator Eq.(\ref{eq:spower}), although this is computationally very
cumbersome. Again, we can take advantage of the relative computational ease of
binned autocorrelations here.% 
\footnote{
   Thanks to Bruce Gluckman for pointing this out to us.}
Introducing a normalisation constant $\alpha$, we can write $C(\tau)=\alpha\int
\!dt\, s(t)s(t-\tau) = \alpha\int\! dt\, \sum_{i,j=1}^N
\delta(t-t_i)\delta(t-\tau-t_j)$.  Then again, the binned autocorrelation
function is defined by
$C_{\Delta}(\tau)=(1/\Delta)\int_{\tau-\Delta}^{\tau}d\tau' C(\tau')$.  Now we
carry out both integrals and thus eliminate both delta functions.  If we choose
$\alpha$ such that $C(0)=1$, we obtain:
\be\label{eq:cspike}
   C_{\Delta}(\tau)={ |{\cal B}_{ij}(\tau-\Delta,\tau)|\over N\Delta}
\,.\ee
Thus, all we have to do is to count all possible intervals $t_i-t_j$ in a bin.
The upper panel in Fig.~\ref{fig:rr} shows the binned autocorrelation function
with bin size $\Delta =0.02$~sec up to a lag of 6~sec for the heart beat data
shown in Fig.~\ref{fig:rrseries}. Superimposed is the corresponding curve for a
surrogate that has been generated with the deviation from the binned
autocorrelations of the data as a cost function.  The two curves are
practically indistinguishable. In this particular case, most of the structure
is given by the mean periodicity of the data. The lower trace of the same
figure shows that even a random scramble shows very similar (but not identical)
correlations. Information about the main periodicity is already contained in
the distribution of inter-beat intervals which is preserved under permutation.

As with unevenly sampled data, the choice of binning and the maximal lag are
somewhat delicate and not that much practical experience exists. It is
certainly again recommendable to avoid empty bins. The possibility to limit the
temporal range of $C_{\Delta}(\tau)$ is a powerful instrument to keep
computation time within reasonable limits.

\section{Questions of interpretation}\label{sec:interpret}
Having set up all the ingredients for a statistical hypothesis test of
nonlinearity, we may ask what we can learn from the outcome of such a test.
The formal answer is of course that we have, or have not, rejected a specific
hypothesis at a given level of significance. How interesting this information
is, however, depends on the null hypothesis we have chosen. The test is most
meaningful if the null hypothesis is plausible enough so that we are prepared
to believe it in the lack of evidence against it. If this is not the case, we
may be tempted to go beyond what is justified by the test in our
interpretation. Take as a simple example a recording of hormone concentration
in a human. We can test for the null hypothesis of a stationary Gaussian linear
random process by comparing the data to phase randomised Fourier surrogates.
Without any test, we know that the hypothesis cannot be true since hormone
concentration, unlike Gaussian variates, is strictly non-negative.  If we
failed to reject the null hypothesis by a statistical argument, we will
therefore go ahead and reject it anyway by common sense, and the test was
pointless.  If we did reject the null hypothesis by finding a coarse-grained
``dimension'' which is significantly lower in the data than in the surrogates,
the result formally does not give any new information but we might be tempted
to speculate on the possible interpretation of the ``nonlinearity'' detected.

This example is maybe too obvious, it was meant only to illustrate that the
hypothesis we test against is often not what we would actually accept to be
true. Other, less obvious and more common, examples include signals which are
known (or found by inspection) to be non-stationary (which is not covered by
most null hypotheses), or signals which are likely to be measured in a static
nonlinear, but non-invertible way. Before we discuss these two specific caveats
in some more detail, let us illustrate the delicacy of these questions with a
real data example.

\bfig%
\centerline{% 
   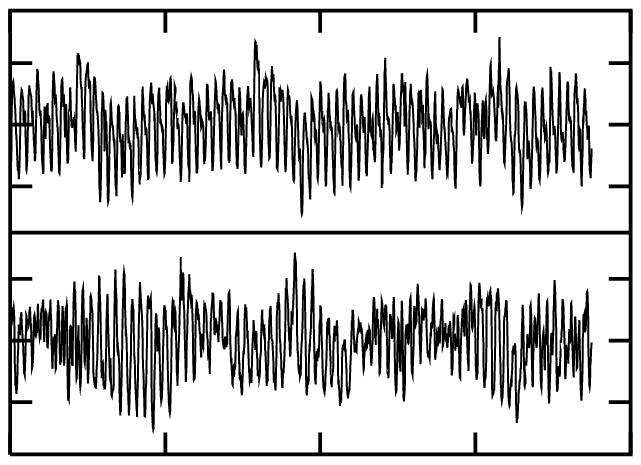% 
}
\caption[]{\label{fig:seizure}Intracranial neuronal potential recording during
  an epileptic seizure (upper) and a surrogate data set with the same linear
  correlations and the same amplitude distribution (lower). The data was kindly
  provided by K. Lehnertz and C. Elger.}  
\end{figure}

\bfig%
\centerline{% 
   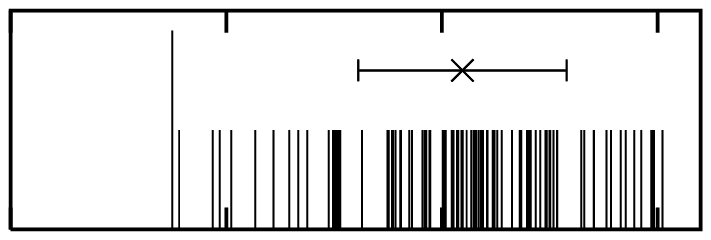% 
}
   \caption[]{\label{fig:barcode}
   Surrogate data test for the data shown in Fig.\ref{fig:seizure}. Since the
   prediction error is lower for the data (longer line) than for 99 surrogates
   (shorter lines), the null hypothesis may be rejected at the 99\%  level of
   significance. The error bar indicates the mean and standard deviation of the
   statistic computed for the surrogates.}
\end{figure}

Figure~\ref{fig:seizure} shows as an intra-cranial recording of the neuronal
electric field during an epileptic seizure, together with one iteratively
generated surrogate data set~\cite{surrowe} that has the same amplitude
distribution and the same linear correlations or frequency content as the data.
We have eliminated the end-point mismatch by truncating the series to 1875
samples. A test was scheduled at the 99\%  level of significance, using
nonlinear prediction errors (see Eq.(\ref{eq:error}), $m=3$, $\tau=5$,
$\epsilon=0.2$) as a discriminating statistics. The nonlinear correlations we
are looking for should enhance predictability and we can thus perform a
one-sided test for a significantly {\em smaller} error. In a test with one data
set and 99 surrogates, the likelihood that the data would yield the smallest
error by mere coincidence is exactly 1 in 100. Indeed, as can be seen in
Fig.~\ref{fig:barcode}, the test just rejects the null hypothesis.

\bfig%
\centerline{% 
   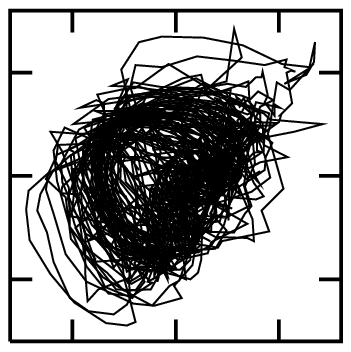%
   \hspace*{-0.7cm}%
   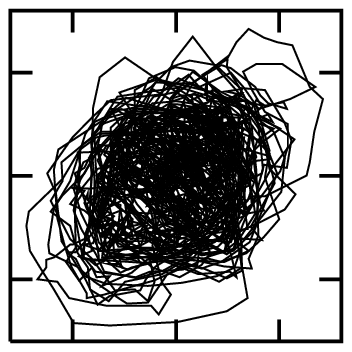% 
}
\caption[]{\label{fig:seizdel}
   Left: Same data set as in Fig.~\ref{fig:seizure}, rendered in time delay
   coordinates. Right: A surrogate data set plotted in the same way.}
\end{figure}

Unfortunately, the test itself does not give any guidance as to what kind of
nonlinearity is present and we have to face notoriously ill-defined questions
like what is the most {\em natural} interpretation.  Similar spike-and-wave
dynamics as in the present example has been previously reported~\cite{FEEG} as
chaotic, but these findings have been questioned~\cite{TEEG}.
Hern\'andez and coworkers~\cite{cuba} have suggested a stochastic limit cycle
as a simple way of generating spike-and-wave-like dynamics.

If we represent the data in time delay coordinates --- which is what we would
usually do with chaotic systems --- the nonlinearity is reflected by the
``hole'' in the centre (left panel in Fig.~\ref{fig:seizdel}). A linear
stochastic process could equally well show oscillations, but its amplitude
would fluctuate in a different way, as we can see in the right panel of the
same figure for an iso-spectral surrogate. It is difficult to answer the
question if the nonlinearity could have been generated by a static mechanism
like the measurement process (beyond the invertible rescaling allowed by the
null hypothesis). Deterministic chaos in a narrower sense seems
rather unlikely if we regard the prediction errors shown in
Fig.~\ref{fig:barcode}: Although significantly lower than that of the
surrogates, the absolute value of the nonlinear prediction error is still more
than 50\%  of the rms amplitude of the data (which had been rescaled to unit
variance). Not surprisingly, the correlation integral (not shown here) does not
show any proper scaling region either. Thus, all we can hand back to the
clinical researchers is a solid statistical result but the insight into what
process {\em is} generating the oscillations is limited.

A recent suggestion for surrogates for the validation of {\em unstable periodic
orbits} (UPOs) may serve as an example for the difficulty in interpreting
results for more fancy null hypothesis. Dolan and coworkers~\cite{witt}
coarse-grain amplitude adjusted data in order to extract a transfer matrix that
can be iterated to yield typical realisations of a Markov chain.% 
\footnote{ 
   Contrary to what is said in Ref.~\cite{witt}, binning a two dimensional
   distribution yields a first order (rather than a second order) Markov
   process, for which a three dimensional binning would be needed to include
   the image distribution as well.}
The rationale there is to test if the finding of a certain number of UPOs could
be coincidental, that is, not generated by dynamical structure.  Testing
against an order $D$ Markov model removes dynamical structure beyond the
``attractor shape'' (AS) in $D+1$ dimensions. It is not clear to us what the
interpretation of such a test would be. In the case of a rejection, they would
infer a dynamical nature of the UPOs found. But that would most probably mean
that in some higher dimensional space, the dynamics could be successfully
approximated by a Markov chain acting on a sufficiently fine mesh. This is at
least true for finite dimensional dynamical systems. In other words, we cannot
see what sort of dynamical structure would generate UPOs but not show its
signature in some higher order Markov approximation.

\subsection{Non-dynamic nonlinearity}\label{sec:rev}
A non-invertible measurement function is with current methods indistinguishable
from dynamic nonlinearity. The most common case is that the data are squared
moduli of some underlying dynamical variable. This is supposed to be true for
the celebrated sunspot numbers. Sunspot activity is generally connected with
magnetic fields and is to first approximation proportional to the squared field
strength.  Obviously, sunspot numbers are non-negative, but also the null
hypothesis of a monotonically rescaled Gaussian linear random process is to be
rejected since taking squares is not an invertible operation. Unfortunately,
the framework of surrogate data does not currently provide a method to test
against null hypothesis involving noninvertible measurement functions. Yet
another example is given by linearly filtered time series.  Even if the null
hypothesis of a monotonically rescaled Gaussian linear random process is true
for the underlying signal, it is usually not true for filtered copies of it, in
particular sequences of first differences, see Prichard~\cite{dean} for a
discussion of this problem.

The catch is that nonlinear deterministic dynamical systems may produce
irregular time evolution, or {\em chaos}, and the signals generated by such
processes will be easily found to be nonlinear by statistical methods. But many
authors have confused cause and effect in this logic: deterministic chaos does
imply nonlinearity, but not vice versa. The confusion is partly due to the
heavy use of methods inspired by chaos theory, leading to arguments like ``If
the fractal dimension algorithm has power to detect nonlinearity, the data must
have a fractal attractor!'' Let us give a very simple and commonplace example
where such a reasoning would lead the wrong way.

One of the most powerful~\cite{power,theiler1,diks2} indicators of nonlinearity
in a time series is the change of statistical properties introduced by a
reversal of the time direction: Linear stochastic processes are fully
characterised by their power spectrum which does not contain any information on
the direction of time.  One of the simplest ways to measure time asymmetry is
by taking the first differences of the series to some power, see
Eq.(\ref{eq:skew}).  Despite its high discrimination power, also for many but
not all dynamical nonlinearities, this statistic has not been very popular in
recent studies, probably since it is rather unspecific about the nature of the
nonlinearity. Let us illustrate this apparent flaw by an example where time
reversal asymmetry is generated by the measurement process.

\bfig%
\centerline{% 
   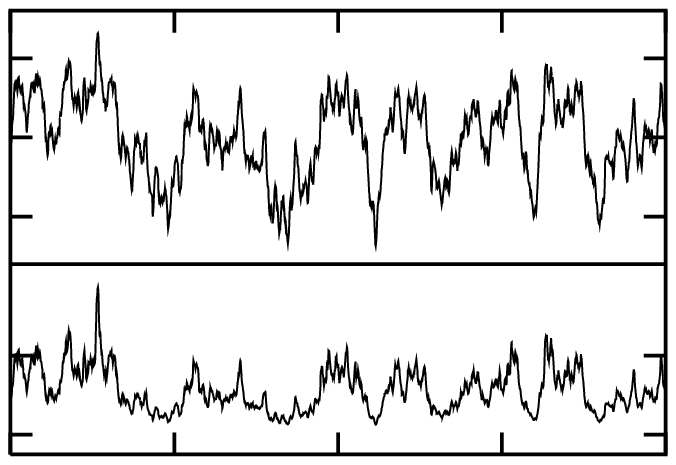% 
}
\caption[]{\label{fig:ar2}Upper panel: Output of the linear autoregressive
   process $x_n=1.6x_{n-1}-0.61x_{n-2}+\eta_n$. Lower panel: the same after 
   monotonic rescaling by $s_n=e^{x_n/2}$.}
\end{figure}
 
\bfig%
\centerline{% 
   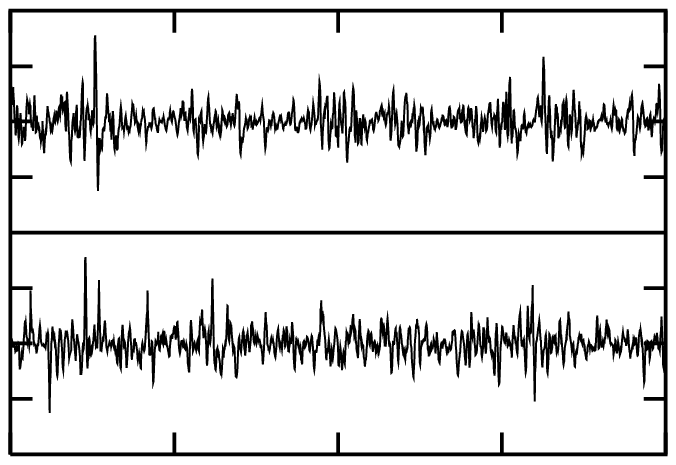% 
}
\caption[]{\label{fig:asym}Moving differences $s_n-s_{n-5}$ of the sequence
   shown in Fig.~\ref{fig:ar2} (upper), and a surrogate time series (lower).  A
   formal test shows that the nonlinearity is significant at the 99\%  level.}
\end{figure}

Consider a signal generated by a second order autoregressive (AR(2)) process
$x_n=1.6x_{n-1}-0.61x_{n-2}+\eta_n$. The sequence $\{\eta_n\}$ consists of
independent Gaussian random numbers with a variance chosen such that the data
have unit variance.  A typical output of 2000 samples is shown as the upper
panel in Fig.~\ref{fig:ar2}. Let the measurement be such that the data is
rescaled by the strictly monotonic function $s_n=e^{x_n/2}$, The resulting
sequence (see the lower panel in Fig.~\ref{fig:ar2}) still satisfies the null
hypothesis formulated above. This is no longer the case if we take differences
of this signal, a linear operation that superficially seems harmless for a
``linear'' signal. Taking differences turns the up-down-asymmetry of the data
into a forward-backward asymmetry. As it has been pointed out by
Prichard,\cite{dean} the static nonlinearity and linear filtering are not
interchangeable with respect to the null hypothesis and the sequence
$\{z_n=s_n-s_{n-5}=e^{x_n/2} - e^{x_{n-5}/2}\}$ must be considered nonlinear in
the sense that it violates the null hypothesis.  Indeed, such a sequence (see
the upper panel in Fig.~\ref{fig:asym}) is found to be nonlinear at the 99\% 
level of significance using the statistics given in Eq.(\ref{eq:skew}), but
also using nonlinear prediction errors. (Note that the nature of the statistic
Eq.(\ref{eq:skew}) requires a two-sided test.) A single surrogate series is
shown in the lower panel of Fig.~\ref{fig:asym}. The tendency of the data to
raise slowly but to fall fast is removed in the linear surrogate, as it should.

\bfig%
\centerline{% 
   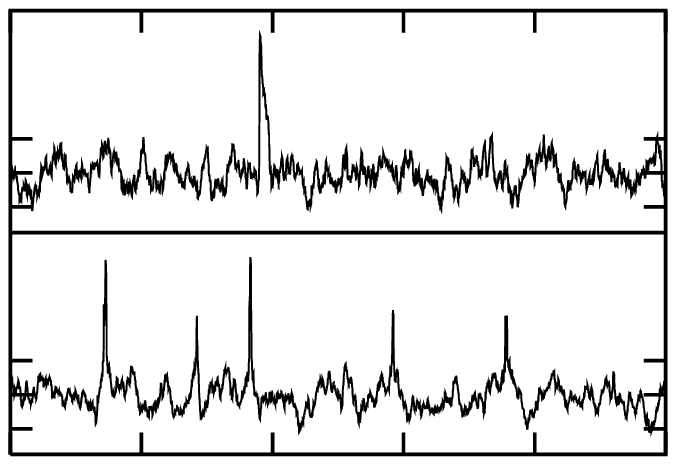% 
}
\caption[]{\label{fig:spike}A single spike is artificially introduced in an
   otherwise linear stochastic time sequence (upper). In the surrogate time
   series (lower), this leads to multiple short spikes. Although the surrogate
   data has the same frequency content and takes on the same set of values as
   the data, the remnants of the spike will lead to the
   detection of nonlinearity.} 
\end{figure}

\subsection{Non-stationarity}\label{sec:nonstat}
It is quite common in bio-medical time series (and elsewhere) that otherwise
harmless looking data once in a while are interrupted by a singular event, for
example a spike.  It is now debatable whether such spikes can be generated by a
linear process by nonlinear rescaling.  We do not want to enter such a
discussion here but merely state that a time series that covers only one or a
few such events is not suitable for the statistical study of the spike
generation process. The best working assumption is that the spike comes in by
some external process, thus rendering the time series non-stationary. In any
case, the null hypotheses we are usually testing against are not likely to
generate such singular events autonomously.  Thus, typically, a series with a
single spike will be found to violate the null hypothesis, but, arguably, the
cause is non-stationarity rather than non-linearity. Let us discuss as a simple
example the same AR(2) process considered previously, this time without any
rescaling. Only at a single instant, $n=1900$, the system is kicked by a large
impulse instead of the Gaussian variate $\eta_{1900}$. This impulse leads to
the formation of a rather large spike.  Such a sequence is shown in
Fig.~\ref{fig:spike}. Note that due to the correlations in the process, the
spike covers more than a single measurement.

When we generate surrogate data, the first observation we make is that it takes
the algorithm more than 400 iterations in order to converge to a reasonable
tradeoff between the correct spectrum and the required distribution of
points. Nevertheless, the accuracy is quite good --- the spectrum is correct
within 0.1\%  of the rms amplitude. Visual inspection of the lower panel of
Fig.~\ref{fig:spike} shows that the spectral content --- and the assumed values
--- during the single spike are represented in the surrogates by a large number
of shorter spikes. The surrogates cannot know of an external kick. The visual
result can be confirmed by a statistical test with several surrogates, equally
well (99\%  significance) by a time asymmetry statistic or a nonlinear
prediction error.

If non-stationarity is known to be present, it is necessary to include it in
the null hypothesis explicitly. This is in general very difficult but can be
undertaken in some well behaved cases. In Sec.~\ref{sec:including} we discussed
the simplest situation of a slow drift in the calibration of the data. It has
been shown empirically~\cite{poster} that a slow drift in system parameters is
not as harmful as expected~\cite{Timmer2}. It is possible to generate
surrogates for sliding windows and restrict the discriminating statistics to
exclude the points at the window boundaries. It is quite obvious that special
care has to be taken in such an analysis.

\section[Conclusions: Testing a Hypothesis vs.\\ Testing Against Surrogates]% 
  {Conclusions: Testing a Hypothesis\\ vs. Testing Against Surrogates} 
Most of what we have to say about the interpretation of surrogate data tests,
and spurious claims in the literature, can be summarised by stating that there
is no such thing in statistics as testing a result {\em against
surrogates}. All we can do is to test a null hypothesis. This is more than a
difference in words. In the former case, we assume a result to be true unless
it is rendered obsolete by finding the same with trivial data. In the latter
case, the only one that is statistically meaningful, we assume a more or less
trivial null hypothesis to be true, unless we can reject it by finding
significant structure in the data.

As everywhere in science, we are applying Occam's razor: We seek the simplest
--- or least interesting --- model that is consistent with the data. Of course,
as always when such categories are invoked, we can debate what is
``interesting''. Is a linear model with several coefficients more or less
parsimonious than a nonlinear dynamical system written down as a one line
formula? People unfamiliar with spectral time series methods often find their
use and interpretation at least as demanding as the computation of correlation
dimensions. From such a point of view it is quite natural to take the
nonlinearity of the world for granted, while linearity needs to be established
by a test {\em against surrogates}.

The reluctance to take surrogate data as what they are, a means to test a null
hypothesis, is partly explainable by the choice of null hypotheses which are
currently available for proper statistical testing.  As we have tried to
illustrate in this paper, recent efforts on the generalisation of randomisation
schemes broaden the repertoire of null hypotheses. The hope is that we can
eventually choose one that is general enough to be acceptable if we fail to
reject it with the methods we have. Still, we cannot prove that there is no
dynamics in the process beyond what is covered by the null hypothesis.  From a
practical point of view, however, there is not much of a difference between
structure that is not there and structure that is undetectable with our
observational means.

\section*{Acknowledgements}
Most data sets shown in this paper are publicly available and the sources were
cited in the text. Apart from these, K.\ Lehnertz and C. Elger at the
University clinic in Bonn kindly permitted us to use their epilepsy
data. Thomas Sch\"urmann at the WGZ-Bank D\"usseldorf is acknowledged for
providing financial time series data. A fair fraction of the ideas and opinions
expressed in this paper we had the opportunity to discuss intensively over the
past few years with a number of people, most notably James Theiler, Danny
Kaplan, Dimitris Kugiumtzis, Steven Schiff, Floris Takens, Holger Kantz, and
Peter Grassberger. We thank Michael Rosenblum for explaining to us the spectral
estimation of spike trains. Bruce Gluckman pointed out the computational
advantage of binned correlations over spike train spectra. Finally, we
acknowledge financial support by the SFB 237 of the Deutsche
Forschungsgemeinschaft.

\Appendix
\section{The TISEAN implementation}\label{app:A}
Starting with the publication of source code for a few nonlinear time series
algorithms by Kantz and Schreiber~\cite{ourbook}, a growing number of programs
has been put together to provide researchers with a library of common tools.
The TISEAN software package is freely available in source code form and an
introduction to the contained methods has been published in
Ref.~\cite{tisean}. More recent versions of the package ($\ge 2.0$) contain a
comprehensive range of routines for the generation and testing of surrogate
data. The general constrained randomisation scheme described in
Sec.~\ref{sec:anneal} is implemented as an extendable framework that allows for
the addition of further cost functions with relatively little effort. With few
exceptions, all the code used in the examples in this paper is publicly
available as part of TISEAN~2.0.

\subsection{Measures of nonlinearity}
A few programs in the package directly issue scalar quantities that can be used
in nonlinearity testing. These are the zeroth order nonlinear predictors ({\tt
predict} and {\tt zeroth}) which implement Eq.(\ref{eq:error}) and the time
reversibility statistic ({\tt timerev}) implementing Eq.(\ref{eq:skew}).  For a
couple of other quantities, we have deliberately omitted a black box algorithm
to turn the raw results into a single number. A typical example are the
programs for dimension estimation ({\tt d2}, {\tt c2}, {\tt c2naive}, and {\tt
c1}) which compute correlation sums for ranges of length scales $\epsilon$ and
embedding dimensions $m$. For dimension estimation, these curves have to be
interpreted with due care to establish scaling behaviour and convergence with
increasing $m$. Single numbers issued by black box routines have lead to too
many spurious results in the literature. Researchers often forget that such
numbers are not interpretable as fractal dimensions at all but only useful for
comparison and classification. Without genuine scaling at small length scales,
a data set that gives $\hat{D}_2=4.2$ by some ad hoc method to estimate
$\hat{D}_2$ cannot be said to have more degrees of freedom, or be more
``complex'' than one that yields $\hat{D}_2=3.5$.

This said, users are welcome to write their own code to turn correlation
integrals, local slopes ({\tt c2d}), Takens' estimator ({\tt c2t}), or Gaussian
Kernel correlation integrals ({\tt c2g}) into nonlinearity measures.  The same
situation is found for Lyapunov exponents ({\tt lyap\_k}, {\tt lyap\_r}),
entropies ({\tt boxcount}) and other quantities. Since all of these have
already been described in Ref.~\cite{tisean}, we refer the reader there for
further details.

\subsection{Iterative FFT surrogates} 
The workhorse for the generation of surrogate data within the TISEAN package is
the program {\tt surrogates}. It implements the iterative Fourier based scheme
introduced in Ref.~\cite{surrowe} and discussed in Sec.~\ref{sec:iterative}. It
has been extended to be able to handle multivariate data as discussed in
Sec.~\ref{sec:multi1}.  An FFT routine is used that can handle data sets of $N$
points if $N$ can be factorised using prime factors 2, 3, and 5 only.  Routines
that take arbitrary $N$ will end up doing a slow Fourier transform if $N$ is
not factorisable with small factors.  Occasionally, the length restriction
results in the loss of a few points. 

The routine starts with a random scramble as $\{\overline{r}_n^{(0)}\}$,
performs as many iterates as necessary to reach a fixed point and then prints
out $\overline{r}_n^{(\infty)}$ or $\overline{s}_n^{(\infty)}$, as
desired. Further, the number of iterations is shown and the residual root mean
squared discrepancy between $\overline{r}_n^{(\infty)}$ and
$\overline{s}_n^{(\infty)}$.  The number of iterations can be limited by an
option. In particular, $i=0$ gives the initial scramble as
$\{\overline{r}_n^{(0)}\}$ or a non-rescaled FFT surrogate as
$\{\overline{s}_n^{(0)}\}$. The first iterate, $\{\overline{r}_n^{(1)}\}$, is
approximately (but not quite) equivalent to an AAFT surrogate. It is advisable
to evaluate the residual discrepancy whenever the algorithm took more than a
few iterations. In cases of doubt if the accuracy is sufficient, it may be
useful to plot the autocorrelation function ({\tt corr}  or {\tt autocor})
of the data and $\overline{r}_n^{(\infty)}$, and, in the multivariate case,
the cross-correlation function ({\tt xcor}) between the channels.
The routine can generate up to 999 surrogates in one call.

Since the periodicity artefact discussed in Sec.~\ref{sec:period} can lead to
spurious test results, we need to select a suitable sub-sequence of the data
before making surrogates. For this purpose, TISEAN contains the program {\tt
endtoend}. Let $\{s^{(n_0)}_n=s_{n+n_0}\}$ be a sub-sequence of length
$\tilde{N}$ and offset $n_0$. The program then computes the contribution of the
end-to-end mismatch $(s^{(n_0)}_1-s^{(n_0)}_{\tilde{N}})^2$ to the total power
in the sub-sequence:
\be
   \gamma_{\rm\scriptsize jump}^{(\tilde{N},n_0)} = 
        {(s^{(n_0)}_1-s^{(n_0)}_{\tilde{N}})^2 \over 
         \sum_{n=1}^{\tilde{N}} 
        (s^{(n_0)}_n-\av{s^{(n_0)}_{\rule{0pt}{1ex}}})^2}
\ee
as well as the contribution of the mismatch in the first derivative
\be
   \gamma_{\rm\scriptsize slip}^{(\tilde{N},n_0)} = 
        {[(s^{(n_0)}_2-s^{(n_0)}_1) -
          (s^{(n_0)}_{\tilde{N}}-s^{(n_0)}_{\tilde{N}-1})]^2 \over 
         \sum_{n=1}^{\tilde{N}} 
         (s^{(n_0)}_n-\av{s^{(n_0)}_{\rule{0pt}{1ex}}})^2}
\ee
and the weighted average 
\be
   \gamma^{(\tilde{N},n_0)} = 
        w\; \gamma_{\rm\scriptsize jump}^{(\tilde{N},n_0)} 
      + (1-w)\; \gamma_{\rm\scriptsize slip}^{(\tilde{N},n_0)}
\,.\ee 
The weight $w$ can be selected by the user and is set to 0.5 by default.
For multivariate data with $M$ channels, $(1/M)\sum_{m=1}^M
\gamma^{(\tilde{N},n_0)}_m$ is used. 

Now the program goes through a sequence of decreasing $\tilde{N}=2^i3^j5^k,
\quad i,j,k\in {\cal N}$, and for each $\tilde{N}$ determines $n_0^*$ such that
$\gamma^{(\tilde{N},n_0^*)}$ is minimal. The values of $\tilde{N}$, $n_0^*$,
and $\gamma^{(\tilde{N},n_0^*)}$ are printed whenever $\gamma$ has
decreased. One can thus easily find a sub-sequence that achieves negligible end
point mismatch with the minimal loss of data.

\subsection{Annealed surrogates} 
For cases where the iterative scheme does not reach the necessary accuracy, or
whenever a more general null hypothesis is considered, the TISEAN package
offers an implementation of the constrained randomisation algorithm using a
cost function minimised by simulated annealing, as introduced in
Ref.~\cite{anneal} and described in Sec.~\ref{sec:anneal}. Since one of the
main advantages of the approach is its flexibility, the implementation more
resembles a toolbox than a single program. The main driving routine {\tt
randomize} takes care of the data input and output and operates the simulated
annealing procedure.  It must be linked together with modules that implement a
cooling schedule, a cost function, and a permutation scheme. Within TISEAN,
several choices for each of these are already implemented but it is relatively
easy to add individual variants or completely different cost functions, cooling
or permutation schemes. With the development structure provided, the final
executables will then have names reflecting the components linked together, in
the form {\tt randomize\_}$A$\_$B$\_$C$, where $A$ is a cost function module,
$B$ a cooling scheme, and $C$ a permutation scheme.

Currently, two permutation schemes are implemented. In general, one will use a
scheme {\tt random} that selects a pair at random. It is, however, possible to
specify a list of points to be excluded from the permutations. This is useful
when the time series contains artifacts or some data points are missing and
have been replaced by dummy values. It is planned to add a
temperature-sensitive scheme that selects pairs close in magnitude at low
temperatures.  For certain cost functions (e.g. the spike train spectrum), an
update can only be carried out efficiently if two consecutive points are
exchanged. This is implemented in an alternative permutation scheme {\tt
event}.

The only cooling scheme supported in the present version of TISEAN (2.0) is
exponential cooling ({\tt exp}). This means that whenever a certain condition
is reached, the temperature is multiplied by a factor $\alpha<1$. Apart from
$\alpha$ and the initial temperature $T_0$, two important parameters control
the cooling schedule. Cooling is performed either if a maximal total number of
trials $S_{\rm\scriptsize total}$ is exceeded, or if a maximal number
$S_{\rm\scriptsize succ}$ of trials has been successfull since the last
cooling. Finally, a minimal number of successes $S_{\rm\scriptsize min}$ can be
specified below which the procedure is considered to be ``stuck''. All these
parameters can be specified explicitly.  However, it is sometimes very
difficult to derive reasonable values except by trial and error. Slow cooling
is necessary if the desired accuracy of the constraint is high.  It seems
reasonable to increase $S_{\rm\scriptsize succ}$ and $S_{\rm\scriptsize total}$
with the system size, but also with the number of constraints incorporated in
the cost function. It can be convenient to use an automatic scheme that starts
with fast parameter settings and re-starts the procedure with slower settings
whenever it gets stuck, until a desired accuracy is reached. The initial
temperature can be selected automatically using the following algorithm.  Start
with an arbitrary small initial temperature. Let the system evolve for
$S_{\rm\scriptsize total}$ steps (or $S_{\rm\scriptsize succ}$ successes). If
less than 2/3 of the trials were successes, increase the initial temperature by
a factor of ten to ``melt'' the system. This procedure is repeated until more
than 2/3 successes are reached. This ensures that we start with a temperature
that is high enough to leave all false minima. If the automatic scheme gets
stuck (the low temperature allows too few changes to take place), it re-starts
at the determined melting temperature. At the same time, the cooling rate is
decreased by $\alpha\to\sqrt{\alpha}$, and $S_{\rm\scriptsize total}\to
\sqrt{2}\, S_{\rm\scriptsize total}$.  We suggest to create one surrogate with
the automatic scheme and then use the final values of $T_0$, $\alpha$ and
$S_{\rm\scriptsize total}$ for subsequent runs. Of course, other more
sophisticated cooling schemes may be suitable depending on the specific
situation. The reader is referred to the standard literature~\cite{annealbook}.

Several cost functions are currently implemented in TISEAN. Each of them is of
the general form (\ref{eq:E}) and the constraints can be matched
in either the $L^1$, $L^2$, or the $L^{\infty}$ (or maximum) norms.
In the $L^1$ and $L^2$ norms, autocorrelations are weighted by
$w_{\tau}=1/\tau$ and frequencies by $w_{\omega}=1/\omega$.

Autocorrelations ({\tt auto}, or a periodic version {\tt autop}) are the most
common constraints available. Apart from the type of
average, one has to specify the maximal lag $\tau_{\rm\scriptsize max}$ (see
e.g. Eq.(\ref{eq:cost})). This can save a substantial fraction of the
computation time if only short range correlations are present.  For each
update, only $O(\tau_{\rm\scriptsize max})$ terms have to be updated.

For unevenly sampled data (see Sec.~\ref{sec:uneven}), the cost function {\tt
uneven} implements binned autocorrelations as defined by
Eq.(\ref{eq:cdelta}). The update of the histogram at each annealing step takes
a number of steps proportional to the number of bins. The user has to specify
the bin size $\Delta$ and the total lag time covered contiguously by the bins.

For surrogate spike trains, either the spike train peridogram
Eq.(\ref{eq:spower}) or binned correlations Eq.(\ref{eq:cspike}) can be used.
In the former case, the cost function is coded in {\tt spikespec}.  The user
has to give the total number of frequencies and the frequency
resolution. Internally, the event times $t_n$ are used. A computationally
feasible update is only possible if two consecutive intervals $t_n-t_{n-1}$ and
$t_{n+1}-t_n$ are exchanged by $t_n\to t_{n-1}+t_{n+1}-t_n$ (done by the
permutation scheme {\tt event}).  As a consequence, coverage of permutation
space is quite inefficient. With binned autocorrelations {\tt spikeauto},
intervals are kept internally and any two intervals may be swapped, using the
standard permutation scheme {\tt random}. 

The documentation distributed with the TISEAN package describes how to add
further cost functions. Essentially, one needs to provide cost function
specific option parsing and input/output functions, a module that computes the
full cost function and one that performs an update upon permutation. The latter
should be coded very carefully. First it is the single spot that uses most of
the computation time and second, it must keep the cost function consistent for
all possible permutations. It is advisable to make extensive tests against
freshly computed cost functions before entering production.

In future releases of TISEAN, it is planned to include routines for cross
correlations in mutivariate data, multivariate spike trains, and mixed signals.
We hope that users take the present modular implementation as a starting point
for the implementation of other null hypotheses.

\end{document}